\documentclass[pra,reprint,nofootinbib,showpacs]{revtex4-1}
\usepackage[utf8]{inputenc}
\usepackage[dvipsnames, table]{xcolor}

\usepackage[colorlinks=true]{hyperref}
\usepackage{amsmath}
\hypersetup{ linkcolor = green,           urlcolor  = blue,            citecolor = blue}
\makeatletter
\usepackage{bm}
\usepackage{amsthm} 
\usepackage{graphicx} 

\usepackage{placeins} 

\usepackage{appendix}

\newcommand{\ket}[1]{\left| #1 \right>} 
\let\baraccent=\= 
\renewcommand{\=}[1]{\stackrel{#1}{=}} 

\theoremstyle{definition}

\theoremstyle{remark}

 \def\be{\begin{equation}}
\def\ee{\end{equation}}
\def\bes{\begin{eqnarray}}
\def\ees{\end{eqnarray}}

\begin{document}

\title{Dynamic attenuation scheme in measurement-device-independent quantum key distribution over turbulent channels}

\author{Brian J. Rollick}
\email{brollick@vols.utk.edu}

\affiliation{Department of Physics and Astronomy, University of Tennessee, Knoxville,
	TN 37996-1200, USA.}

\author{Bing Qi}
\email{bingq@cisco.com}
\affiliation{Cisco Systems, San Jose, CA 95134, USA.}

\author{George Siopsis}
\email{siopsis@tennessee.edu}
\affiliation{Department of Physics and Astronomy, University of Tennessee, Knoxville,
	TN 37996-1200, USA.}

\date{\today}

\begin{abstract}
Measurement-device-independent quantum key distribution (MDI QKD) offers great security
in practice because it removes all detector side channels.  However, conducting MDI QKD over  free-space channels is challenging. One of the largest culprits is the mismatched transmittance of the two independent turbulent channels causing a reduced Hong-Ou-Mandel visibility and thus a lower secret key rate. Here we introduce a dynamic attenuation scheme, where the transmittance of each of the two channels is monitored in real time by transmitting bright light pulses from each users to the measurement device. Based on the measured channel transmittance, a suitable amount of attenuation is introduced to the low-loss channel at the measurement device. Our simulation results show a significant improvement of QKD performance, especially when using short raw keys.
\end{abstract} \hspace{10pt}

\maketitle

\section{Introduction} 

Ever since the Bennett-Brassard 1984 (BB84) protocol was proposed, QKD has enjoyed tremendous progress. In particular, free space experiments have progressed from just 30 cm \cite{Bennett1989} to 7600 km satellite-based connections \cite{Liao2017a}.

Despite this progress, numerous hacking techniques have been discovered. While in theory, QKD is secure, realistic implementations deviate from ideal models used in the security proofs. In particular, detectors may be vulnerable to a plethora of attacks, including the detector blinding attack \cite{Makarov2009, Lydersen2011}, time-shift attack \cite{qi2005time}, backflash attack \cite{Pinheiro2018}, and many others (see Jain, \emph{et al.}\ \cite{Jain2016}). 

Two types of countermeasures have been proposed. The first type involves addressing new attacks as they are discovered, adjusting the setup accordingly in order to thwart them. An example of such countermeasures is adding an optical isolator to combat the backflash attack. However, unknown attacks cannot be anticipated. The second type involves device-independent QKD (DI QKD) protocols \cite{acin2007device,barrett2005no,mayers1998quantum}. In this category, a common entanglement source sends photon pairs to Alice and Bob. Because entanglement is monogamous, the protocol is provably secure with the proof relying directly on the violation of Bell's inequalities \cite{Bell1964}. However, a loophole-free Bell test is very challenging in practice \cite{Garg}. For a recent experimental demonstration of DI-QKD, see \cite{zhang2021experimental}.


A more practical protocol, called measurement-device-independent QKD (MDI QKD) \cite{Lo2012}, automatically removes all detector side-channels by employing time-reversed entanglement. In this protocol, Alice and Bob send light pulses to a third party, Charlie, who possesses a Bell-state analyzer based on linear optics and single-photon detection. Charlie projects the input photons to Bell states and publicly announces the measurement results, which allows Alice and Bob to generate a secret key by classical post-processing. MDI QKD has been widely implemented with attenuated laser sources that incorporate the decoy-state protocol \cite{Rubenok2013,tang2014experimental,yin2016measurement}. It has also been implemented on chips \cite{Wei2020} and with cost-effective setups \cite{valivarthi2017cost}. 


The Bell-state analyzer in MDI QKD relies on the Hong-Ou-Mandel (HOM) effect \cite{Hong1987} where photons from Alice and Bob interfere at a 50:50 beam splitter. A high HOM visibility can usually be translated into a low quantum bit error rate (QBER) and therefore a high secret key rate. To achieve a high HOM visibility, photons from Alice and Bob should be indistinguishable in all degrees of freedom. Furthermore, when the MDI QKD is implemented with weak coherent sources, a high HOM visibility requires the average photon numbers from Alice and Bob to be matched at the beam splitter \cite{Wang2017,Moschandreou2018}. 

In practice, the two quantum channels (one from Alice to Charlie, and another from Bob to Charlie) may have different transmittance. One could account for this mismatch by simply adding extra fiber on the low-loss channel so that each channel equally attenuates the light pulses \cite{Rubenok2013}. While being unwieldy in a future quantum network with many users, this approach can improve the HOM visibility at the cost of a lower detection rate. A better solution is the asymmetric MDI QKD protocol where Alice and Bob use different intensity profiles \cite{Xu2013, Wang2019a}. The asymmetric MDI QKD allows Alice and Bob to send different intensities to help account for the asymmetric channel loss. This results in a large improvement over simply adding fiber \cite{Wang2019a}. 

Unfortunately, the above asymmetric MDI QKD protocol requires static channels (such as optical fiber) and may not be applicable in free-space MDI QKD, where the transmittance of each of the two channels fluctuates randomly and independently of the other channel. 
Asymmetric protocols could still help for compensating the mismatch of the average channel losses. To compensate for the channel fluctuation, Alice and Bob would have to know the instantaneous channel transmittance and change their QKD parameters on the scale of milliseconds \cite{Osche2002}.

In free-space BB84 setups, an adaptive post-selection scheme was proposed where a stronger probe beam would be multiplexed with the single photon pulses to monitor the atmospheric transmittance at a given time\cite{Erven_2012,Capraro2012,Vallone2015,Wang2018}. The time blocks with lower transmittance would correspond to a higher QBER. Hence, discarding pulses measured in those blocks could increase the key rate despite reducing the detection rate \cite{Moschandreou2021}. Recent work, such as \cite{Cao2020, Wang2019, Zhu2018}, has also introduced this idea to MDI QKD. 


In this work, we propose a dynamic attenuation scheme to improve the key rate of free-space MDI QKD. Similar to the adaptive post-selection scheme in BB84 QKD, both Alice and Bob transmit strong probe beams with known intensities to Charlie, who determines the channel transmittances in real time by measuring probe beams with classical photo-detectors, and then applies an appropriate amount of attenuation on one of the paths to compensate for the mismatch of channel transmittance. 
The effect is similar to the case of adding extra fiber in asymmetric channels. Our simulations show that using dynamic attenuation makes MDI QKD considerably more robust in turbulence. In the high-turbulence region, our scheme still shows improvement even when we consider a non-zero minimum loss for Charlie's variable attenuators.

Our discussion is organized as follows. Section \ref{sec:2} contains pertinent background for MDI QKD, discusses how turbulence affects transmission, and details our proposed scheme. Section \ref{sec:3} outlines our simulation model and presents our results. Lastly, Section \ref{sec:con} contains a brief discussion of our novel approach and suggests future work. Details of the noise model we used and the secure key calculation are provided in Appendix \ref{keyCalculation}.

\section{Theory}\label{sec:2}

\subsection{Polarization Encoding MDI QKD}

Inspired by time-reversed entanglement \cite{biham1996,inamori2002}, MDI QKD was proposed as a solution to detector side-channel attacks. 
In this protocol, Alice and Bob generate a key by sending laser pulses to a potentially untrusted third party, Charlie, who projects them onto Bell States and publicly announces his results. In general, they may choose time-bin encoding \cite{Ma2012,kaneda2017,liu2013}, phase encoding \cite{pirandola2015}, or polarization encoding \cite{da2013,Tang2014}. Here, we work with polarization encoding.

Phase randomized weak coherent pulses remain common in QKD implementations. To improve the performance of QKD, decoy-state protocols are employed \cite{Hwang2003,Lo2005,wang2005beating}. The original MDI QKD protocol used 3 different intensities \cite{Lo2012,yu2015}. Zhou, \emph{et al.}, then showed a sizeable improvement with a four-intensity method\cite{Zhou2016}. A seven-intensity method was also suggested to account for asymmetric channels in \cite{Wang2019a} where Alice and Bob could choose their signal and decoy parameters independently to account for asymmetric loss. In this work, we only consider channels with identical statistical distribution, so the four-intensity protocol is adopted.

\begin{figure}
    \centering
    \includegraphics[width=0.7\linewidth]{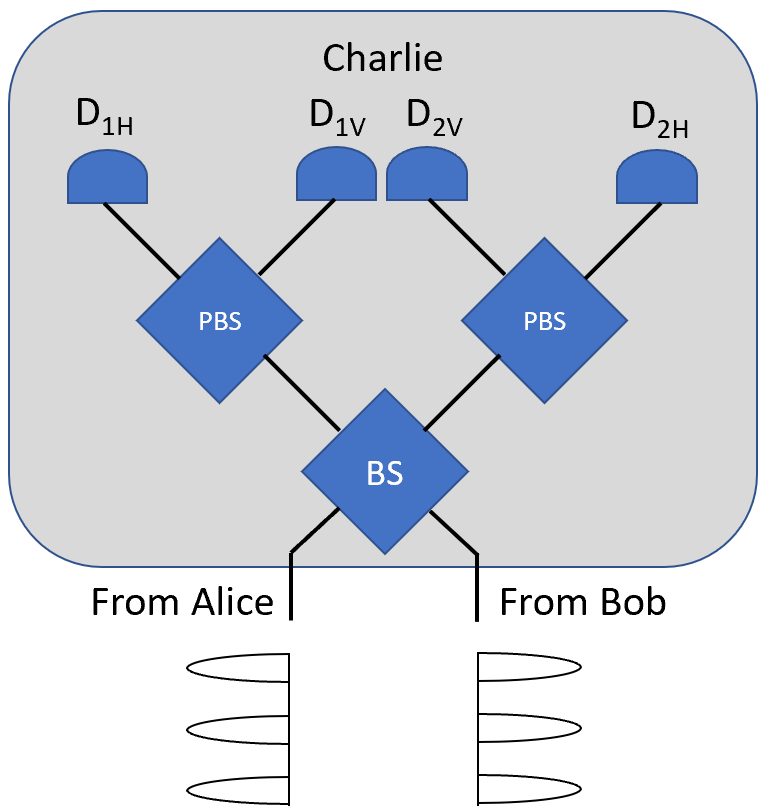}
    \caption{Basic measurement setup for Charlie in a polarization encoding MDI QKD experiment. Alice and Bob send pulsed laser beams to Charlie whose experimental setup consists of a beam splitter (BS), two polarization beam splitters (PBS) and four single-photon detectors (D). Charlie publicly announces his measurement results and one of Alice and Bob may apply bit-flip depending on the Bell state detected and the encoding basis.}
    \label{fig:charlieNormal}
\end{figure}

In MDI QKD with polarization encoding, Alice and Bob encode their random bits on the polarization of weak coherent states, using one of the two bases, rectilinear (Z) or diagonal (X), and Charlie performs Bell-state measurements using a setup depicted in Figure \ref{fig:charlieNormal}. 

A bit of raw key is generated whenever Charlie measures a coincidence of photons with orthogonal polarizations ($V$ and $H$, respectively) using a set of four single-photon detectors, $\{ D_{1H}, D_{1V}, D_{2H}, D_{2V} \}$, and Alice and Bob use the same encoding basis. Charlie announces the outcome $\ket{\psi^-}$ whenever coincidences occur on $D_{1V}D_{2H}$ or $D_{1H}D_{2V}$, and $\ket{\psi^+}$ if coincidences occur on $D_{1H}D_{1V}$ or $D_{2H}D_{2V}$, instead. Other detection patterns are simply discarded.  In the rectilinear basis, errors occur whenever Alice and Bob send the same polarization, and Charlie announces $\ket{\psi^-}$ or $\ket{\psi^+}$. In the diagonal basis, errors occur whenever Alice and Bob send the same polarization and Charlie announces $\ket{\psi^-}$ or Alice and Bob send orthogonal polarization states and Charlie announces $\ket{\psi^+}$.

\begin{figure}
    \centering
    \includegraphics[width=0.7\linewidth]{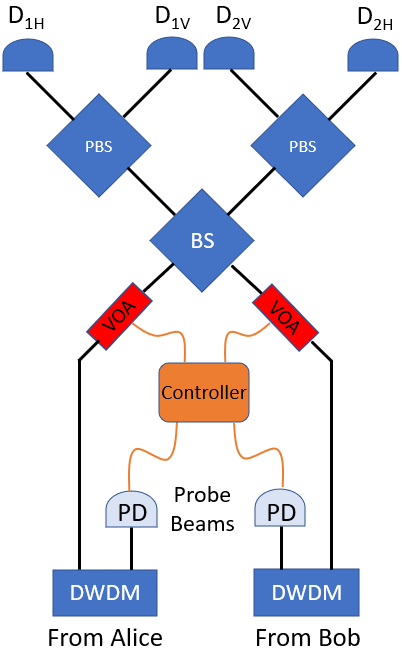}
    \caption{Charlie's measurement setup with dynamic attenuation using probe beams sent by Alice and Bob. The multiplexed probes are separated by DWDMs and their intensities are measured using classical light detectors. Charlie, based on the measurement results of the probe beams, applies a proper amount of attenuation to the channel with lower loss using variable optical attenuators.}
    \label{fig:da_setup}
\end{figure}

Notice in Figure~\ref{fig:charlieNormal} that Charlie only measures in the rectilinear basis. To ensure a low error rate when Alice and Bob use the diagonal basis, Charlie relies on the HOM effect to bunch identical photons from Alice and Bob at the beam splitter. 
High HOM visibility requires Alice's and Bob's photons to be identical in every degree of freedom at the beam splitter. In the case of phase-randomized weak coherent sources, the average photon numbers from Alice and Bob should be matched at the beam splitter \cite{Wang2017, Moschandreou2018}. Consequently, it can be difficult to consistently achieve high HOM visibility in a turbulent atmosphere with fluctuating transmittance. The central idea of our dynamic attenuation scheme it to compensate the transmittance mismatch by dynamically controlling the amount of attenuation introduced.

\subsection{Atmospheric Effects}

The transmittance coefficient of light $\eta$ follows a lognormal distribution through a weak to moderate turbulent channel \cite{Ghassemlooy2012, Goodman2015, Karp2013}. In this regime, we can express the effect of turbulence using two parameters,
the average transmittance $\eta_0$, and the log irradiance variance $\sigma^2$ which characterizes the severity of the
turbulence. The probability distribution of the transmittance coefficient (PDTC) is given by:

\begin{align}
    P(\eta) = \frac{1}{\sqrt{2\pi}\sigma\eta}\textrm{e}^{\frac{-(\textrm{ln}(\frac{\eta}{\eta_0}) + \frac{\sigma^2}{2})^2}{2\sigma^2}}
    \label{Eq:Lognormal}
\end{align}
Very weak to moderately strong turbulence for a 3 km channel has $\sigma^2$ ranging from $10^{-3}$ to about 1.2 at 1550 nm wavelength. After this point, the lognormal distribution for transmittance loses validity \cite{Ghassemlooy2012}. 

The average loss $\eta_0$ can be determined from atmospheric visibility and channel length. Due to the complexity of different atmospheric and aerosol models, software such as MODTRAN \cite{berk1987modtran,Berk2016} and FASCODE \cite{smith1978fascode} are often required to find transmittance for an arbitrary wavelength.

In this work, we consider average losses of $\eta_0 = $ 17 dB, 14 dB, 11 dB, and 8 dB in each channel (excluding the efficiency of detector), and we choose QKD parameters based on a recent free-space MDI-QKD demonstration \cite{Cao2020}. We then simulate the secret key rate using a range of values for $\sigma^2$ and show that dynamic attenuation makes MDI QKD more tolerant of channel fluctuation. 

\subsection{Dynamic Attenuation Scheme} \label{KernelStep}

We propose a scheme where high-speed, low-loss variable optical attenuators (VOA) are placed before Charlie's beam splitter (see Figure \ref{fig:da_setup}) to balance the transmittance fluctuations between the two channels. Both Alice and Bob transmit strong probe beams
with known intensities along the same paths as the QKD signals, but slightly separated in wavelength. Charlie can separate the probe beams from the QKD signals using dense wavelength-division multiplexing (DWDM) technology and determine the channel transmittance in real time by measuring probe beams with classical photo-detectors. He further applies an appropriate amount of attenuation on the high-transmittance path.


The goal of adding additional attenuation is to balance the loss between the two channels and improve the HOM visibility. However, because additional loss could negatively impact the raw key rate, an optimal balance must be found to maximize the final secret key rate, as we discuss below.

If we assume Alice's and Bob's channels are independent, the joint probability distribution is simply:
\begin{equation}\label{eq:PAB}
    P(\eta_A, \eta_B) = P(\eta_A)P(\eta_B),
\end{equation}
where $P(\eta)$ is defined in Eq.\ \eqref{Eq:Lognormal}.

\begin{figure*}[ht!]
    \centering
    \includegraphics[width=0.7\linewidth]{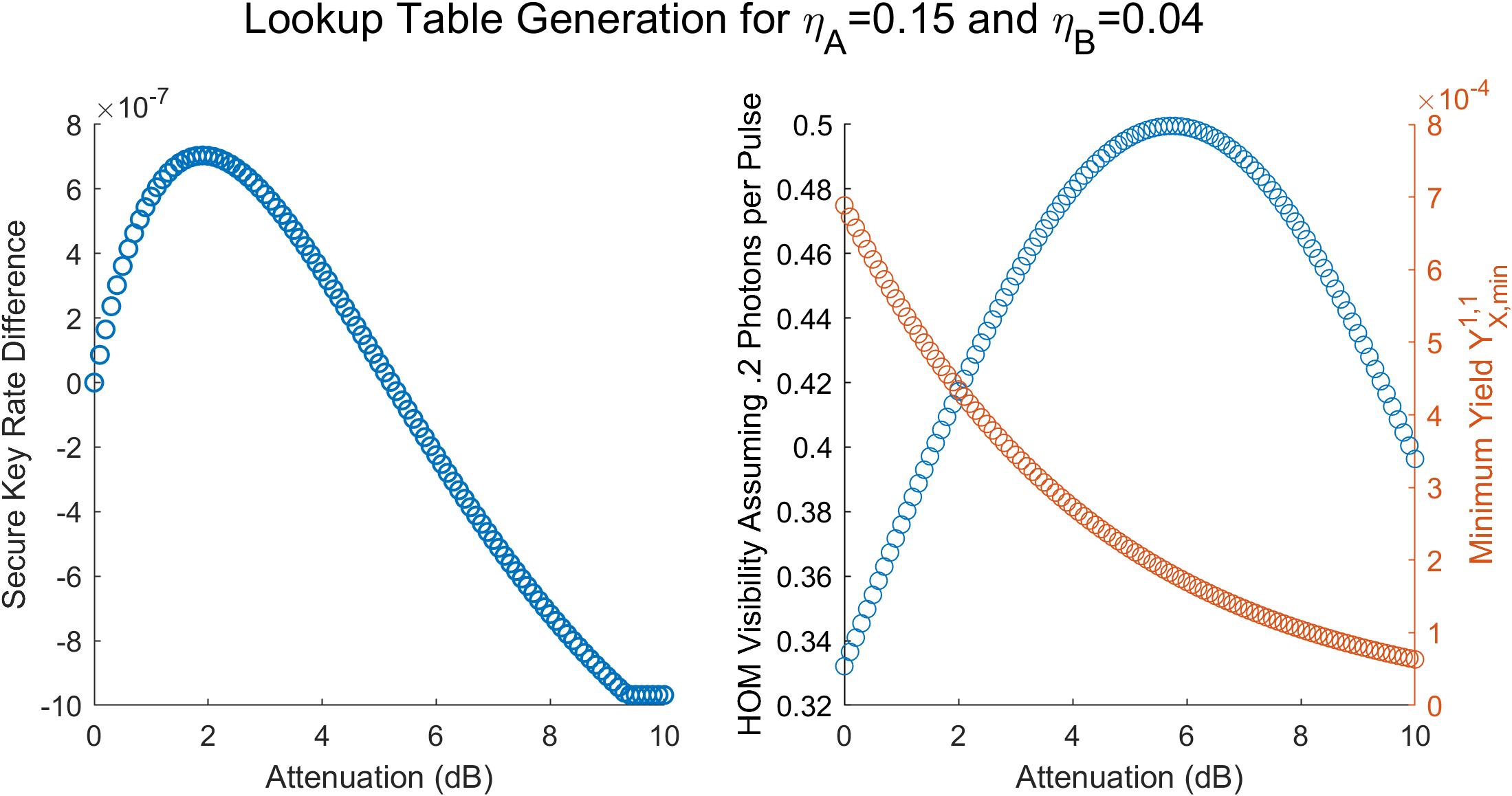}
    \caption{Results using a lookup table for $\eta_A = .15$, $\eta_B = .04$, and $N=10^{13}$ pulses. Left panel: improvement in secure key rate when we attenuate the stronger channel (Alice). Right panel: improvement of the HOM visibility, thereby our estimate of $e_X^{1,1}$, and the decrease in yield as Alice's channel is attenuated.}
    \label{fig:lookupExample}
\end{figure*}

Since the secret key rate is a function of transmittance, one can compute the average key rate using the following integral, given an infinite key length:
\begin{equation}
    R_{ave} = \int_0^1\int_0^1{R(\eta_A, \eta_B)P(\eta_A)P(\eta_B)d\eta_Ad\eta_B},
    \label{eq:keyRateIntegral}
\end{equation}
where we used Eq.\ \eqref{eq:PAB} for the joint probability distribution. $R(\eta_A, \eta_B)$ is the key rate bounded by \cite{Lo2012}:

\begin{align}
    R\geq P_Z^{1,1}Y_Z^{1,1}[1 - H_2(e_X^{1,1})] - Q_Zf_{EC}(E_Z)H_2(E_Z)
    \label{eq:mdiKeyRateAsymp}
\end{align}
where
$P_Z^{1,1}$ is the probability of sending singe-photon pair in the $Z$ basis, $Y_Z^{1,1}$ is the yield of single-photon pair in the $Z$ basis, and $Q_Z$ and $E_Z$ are the gain and error rates, respectively. $e_X^{1,1}$ is the error rate of single photon in the $X$ basis, $f_{EC}$ is the error-correcting efficiency, and $H_2(x)$ is the Shannon binary entropy function.
   
When using dynamic attenuation, 
the joint PDTC is transformed according to 
\begin{equation}\label{eq:kernel}
    P(\eta_A\prime,\eta_B\prime) = \int\int{K(\eta_A\prime,\eta_B\prime,\eta_A,\eta_B)P(\eta_A,\eta_B)d\eta_Ad\eta_B}
\end{equation}
where $K(\eta_A\prime,\eta_B\prime,\eta_A,\eta_B)$ represents the kernel relating the new joint PDTC with the original one.
Because we cannot achieve infinite resolution, a lookup table is used instead of the kernel. 

If we use the lookup table to transform the joint probability distribution, the asymptotic key rate after dynamic attenuation is found from Eq.\ \eqref{eq:keyRateIntegral}. We deduce
\begin{equation}
    R'_{ave}=\int_0^1\int_0^1{R(\eta_A\prime, \eta_B\prime)P(\eta_A\prime,\eta_B\prime)d\eta_A\prime d\eta_B\prime}
    \label{eq:keyRateIntegralPrimed}
\end{equation}
where the integral is evaluated numerically. 

Furthermore, Eq.~\eqref{eq:keyRateIntegralPrimed} must be modified in the case of finite key size, because when additional loss is introduced, the finite size effect is exacerbated. Consequently, rather than integrate the PDTC against the secure key rate, we integrate to find new sifted key and error sizes $\{ n_Z, n_X, m_Z, m_X \}$ and use a bounded version of Eq.~\eqref{eq:mdiKeyRateAsymp} afterward. Thus, we separately compute:
\begin{align}
    n_{X,Z}^{i,j} &= \int_0^1\int_0^1{n_{X,Z}^{i,j}(\eta_A,\eta_B)P(\eta_A\prime,\eta_B\prime)d\eta_A\prime d\eta_B\prime}
    \\m_{X,Z}^{i,j} &= \int_0^1\int_0^1{n_{X,Z}^{i,j}(\eta_A,\eta_B)P(\eta_A\prime,\eta_B\prime)d\eta_A\prime d\eta_B\prime}
    \label{eq:siftedBitIntegral}
\end{align}
where $n_{X,Z}^{i,j}$ and $m_{X,Z}^{i,j}$ represent the number of sifted bits and errors, respectively, in the $X$ and $Z$ bases and for $i, j$ states (signal, or one of the decoy states) from Alice and Bob. Once the sifted bits and errors have been found, we compute the full secure key length using our sifted bits/errors and QKD system parameters. A detailed description of the key calculation can be found in Appendix ~\ref{keyCalculation}.

\section{Parameter Optimization and Kernel Generation}\label{sec:3}
 
In QKD using decoy states \cite{Hwang2003,Lo2005,wang2005beating}, it is essential to optimize the intensity of each state and the probability of sending it. 
Here, we employ the four-intensity protocol, therefore six parameters must be optimized. In particular, Alice and Bob use the set of intensities $\{ s, \mu, \nu, \omega \}$, where $s$ is the signal state intensity, $\mu$ and $\nu$ are the decoy state intensities, and $\omega =0$ is the vacuum state.





We optimize decoy parameters stochastically using MATLAB's built-in genetic algorithm. This is a preferred technique because it requires neither differentiability nor any initial data points. It also runs reasonably quickly on an ultrabook's Ryzen 5 processor.  


Prior to optimization, we choose total number of pulses $N$, $Z$-basis misalignment $e_{dZ}$, $X$-basis misalignment $e_{dX}$, dark counts $Y_0$, detector efficiencies $\eta_D$ and an estimated channel transmittance $\eta_0$. Our choices are taken from a recent free-space MDI-QKD experiment in Ref.\ \cite{Cao2020} and are listed in Table \ref{tab:expParams}.

\begin{table}[ht!]
\centering
\begin{tabular}{||l|l|l|l|l||}
\hline
$\eta_D$ & $e_{dZ}$ & $e_{dX}$ & $f_{EC}$ & $Y_0$  \\
\hline
.5 & .003 & .03 & 1.1 & $7\times10^{-7}$  \\
\hline
\end{tabular}%

\caption{$\eta_D$ is detector efficiency, $e_{dZ}$ and $e_{dX}$ are misalignments in their respective bases, $Y_0$ is the dark count probability and $f_{EC}$ is the error correction efficiency.}
\label{tab:expParams}
\end{table}

Because the lognormal distribution's median is considerably less than the mean for higher turbulence, about 3-6 dB extra loss in each arm must be budgeted into $\eta_0$ at the optimization and lookup stages, compared to the simulation step. We therefore optimize channels assuming $\eta_0 \in \{.01, .02, .04\}$.

In the optimization and lookup table generation, we compute the finite secure key rate using:
\begin{multline}
    R= P_s^2(s^2\textrm{e}^{-2s}Y_{X,min}^{1,1}[1 - H_2(e_X^{1,1,max})] \\- Q_{Z}f_{EC}(E_Z)H_2(E_Z))
    \label{eq:mdiKeyRatefinite}
\end{multline}
where $P_s$ is the probability of sending a signal state, s is the average photon number of the signal state, $Y_{X,min}^{1,1}$ is the lower bound of a  single-photon pair yield in the $X$ basis, and $Q_Z$ and $E_Z$ are the $Z$-basis (signal) gain and error rates, respectively. $e_{X,max}^{1,1}$ is the upper bound of single-photon pair error rate in the $X$ basis, $f_{EC}$ is the error correcting efficiency, and $H_2(x)$ is the Shannon binary entropy function. A detailed calculation can be found in Appendix ~\ref{keyCalculation}.

Once decoy parameters are obtained, the lookup table is generated where the transmittance of Alice's and Bob's channels are varied in the range $0.001\leq\eta_0\leq1$ in increments of about .001. For each pair of transmittances, we compute the key rate using Eq.~\eqref{eq:mdiKeyRatefinite} assuming a static channel. Specifically, we evaluate the key rate after applying an additional .1 dB to the stronger channel until we find the maximum. For example, in Figure \ref{fig:lookupExample} the optimum for a given pair of transmittances occurs near 2 dB of attenuation. Note that although the maximum HOM visibility occurs at about 6 dB, the yield suffers too much to achieve maximum secret key rate. We record the maximum for each pair of transmittances and this is the optimal additional attenuation to be introduced by Charlie using a VOA. 

\begin{figure}[ht!]
	\centering
	\includegraphics[width=1\linewidth]{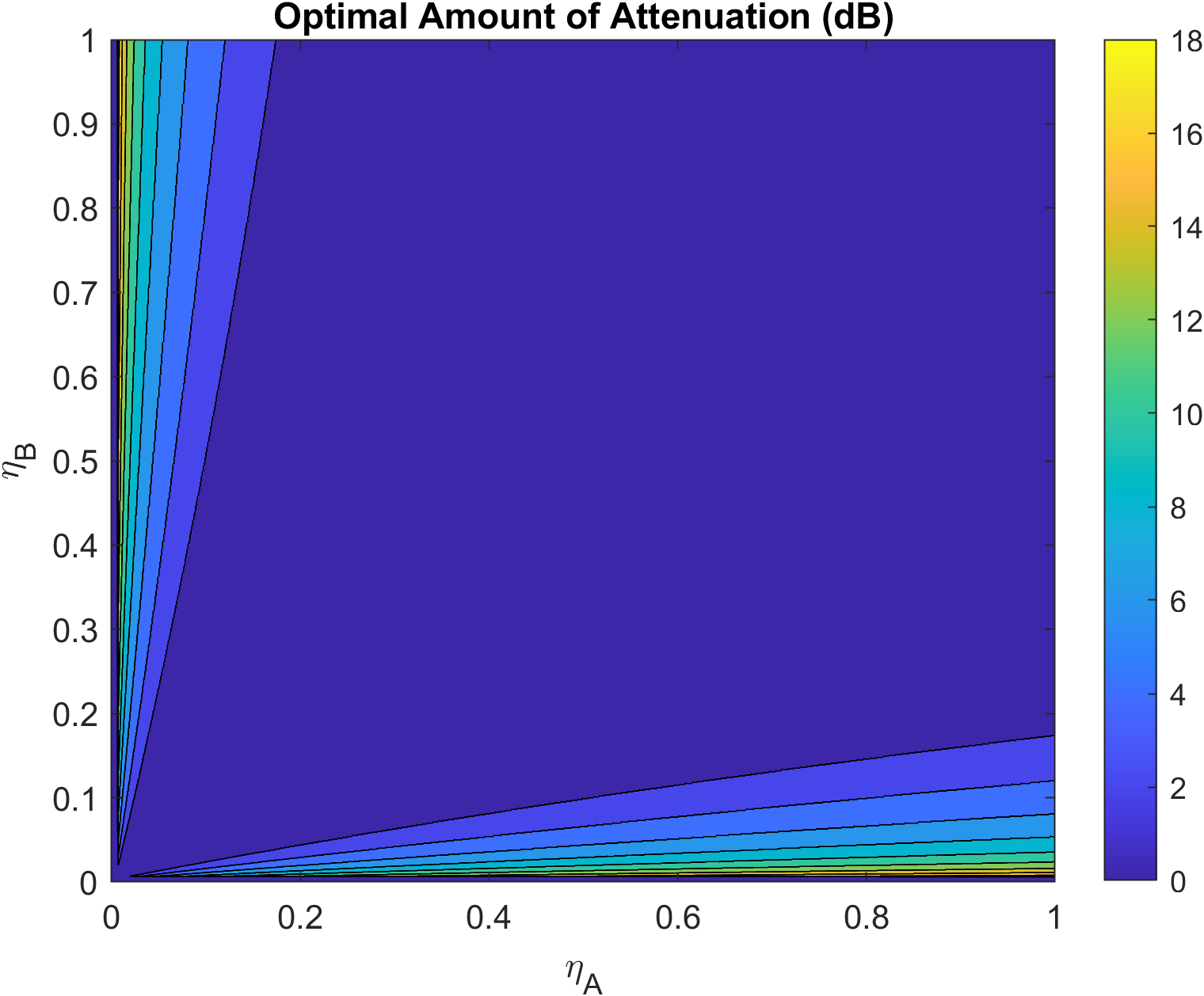}
	\caption{Plot of the optimal amount of attenuation needed to produce the highest key rate for different combinations of Alice's and Bob's transmittances when $N=10^{14}$. The middle of the plot requires no additional attenuation because the transmittances are already close. In high turbulence, Alice's and Bob's channels will likely have very different transmittances and hence dynamic attenuation is useful to enhance the key rate.}
	\label{fig:optimalAttenuation}
\end{figure}

\begin{figure}
	\centering
	\includegraphics[width=1\linewidth]{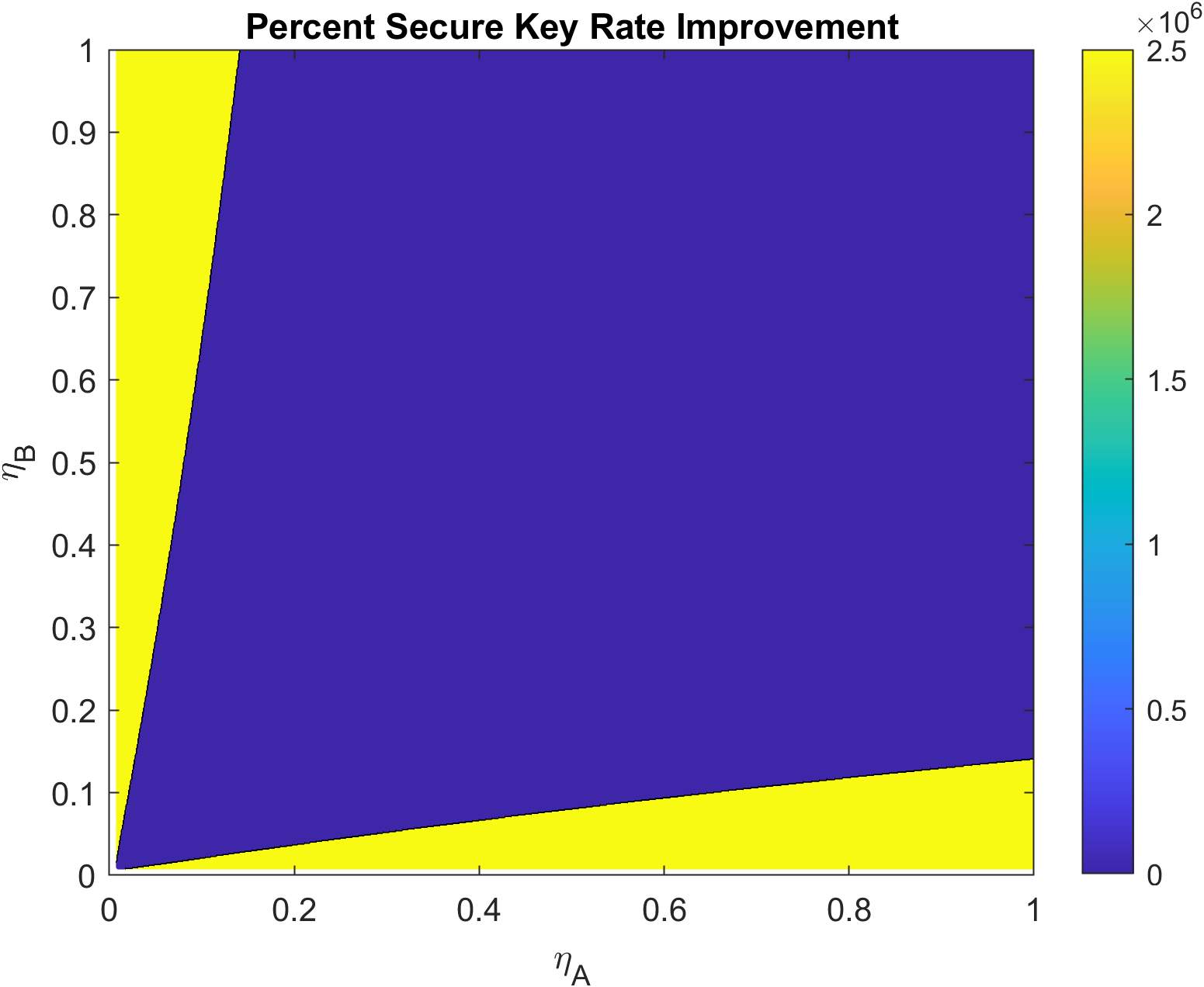}
	\caption{The percent improvement in the secure key rate when the optimal attenuation is applied. In turbulent conditions, when Alice's and Bob's transmittances fluctuate the most, and transmittances are likely very different, Alice and Bob see the greatest benefit of dynamic attenuation.}
	\label{fig:20dBImprovement}
\end{figure}

A plot of the optimal attenuation as a function of transmittance of Alice's and Bob's channels is shown in Figure \ref{fig:optimalAttenuation}. Here, we observe that the optimum is zero additional attenuation for many combinations of transmittances, except when their imbalance is large. However, the channels will most likely be highly imbalanced when the atmospheric turbulence is strong. If we apply the optimal attenuation, the impact on the secret key rate, as shown in Figure \ref{fig:20dBImprovement}, increases precipitously as the imbalance goes up. 

Computing time determines the fineness of the lookup table. The finer the resolution, the more accurately the table will approximate the ideal kernel $K$ (Eq.\ \eqref{eq:kernel}). Improvements can still be seen when the resolution is more coarse than about .001, but the effects are less pronounced.

\begin{figure*}[ht!]
	\centering
	\includegraphics[width=0.8\linewidth]{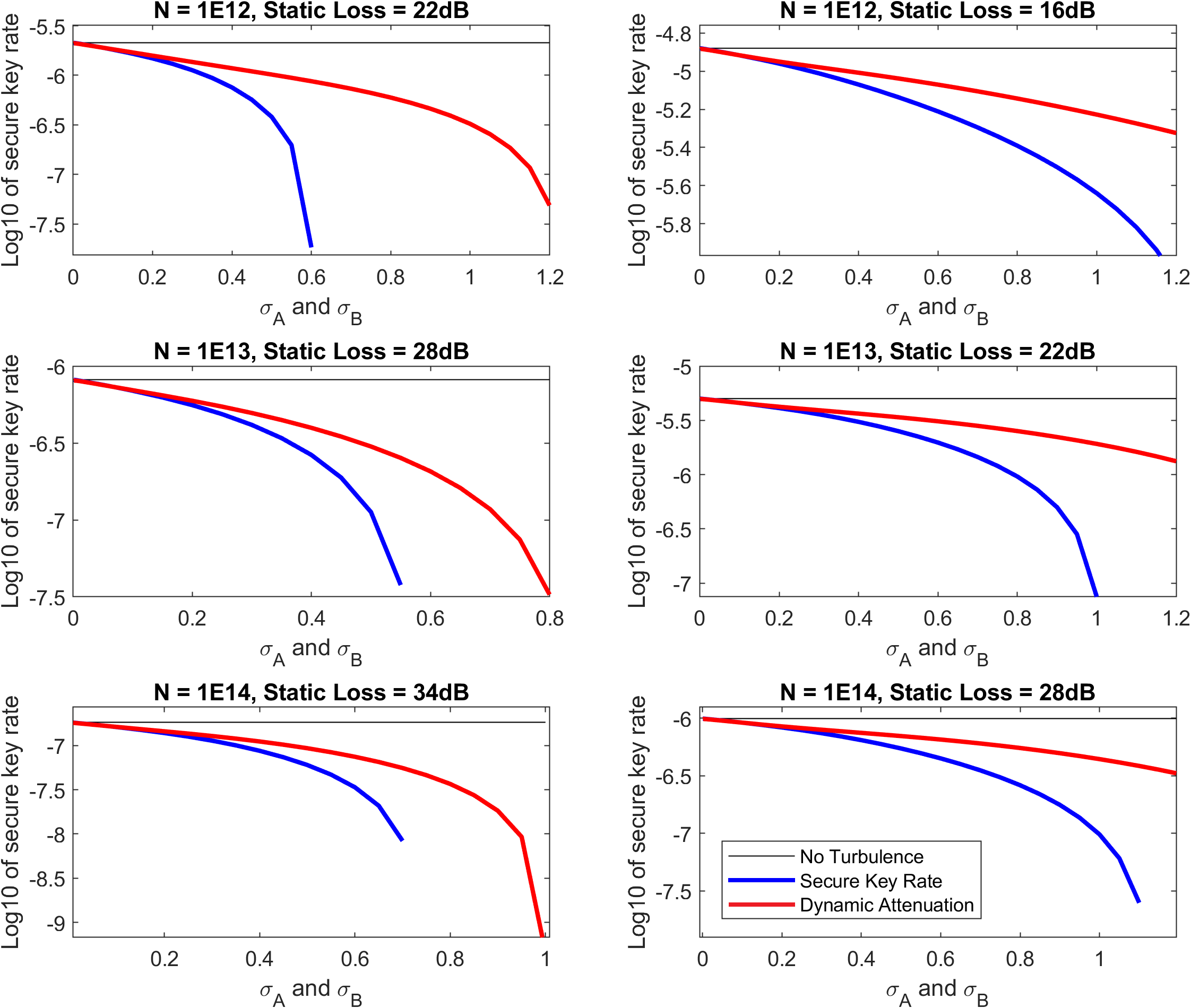}
	\caption{Key rate using dynamic attenuation (red) compared with key rate without dynamic attenuation (blue) vs.\ severity of turbulence for various pulse sizes and average losses. Notice the improvement in the key rate is better for higher turbulence.}
	\label{fig:attachedPlotSigma}
\end{figure*}

\section{Results}

Once the lookup table is produced for a given set of parameters, we produce PDTCs for different atmospheric conditions using Eq.\ \eqref{Eq:Lognormal}. We start with very weak turbulence of $\sigma^2 = .001$ and go up to $\sigma^2 = 1.2$, after which we would need to move beyond the lognormal model of turbulence \cite{Ghassemlooy2012}. Our choices of $\eta_0$ assume losses of 17 dB, 14 dB, 11 dB, and 8 dB in each channel.


To produce each PDTC, we numerically integrate to determine the number of sifted bits for each pair of transmittances, as in Eq.~\eqref{eq:siftedBitIntegral}. The integral is computed by evaluating the PDTCs in .001 increments in the range $0\leq \eta \leq1$, for both Alice's and Bob's channels, and finding the number of sifted bits contributing to the final key. We evaluate the noise model using each pair of transmittances, additionally attenuate the stronger transmittance using the lookup table, and separately determine the number of sifted bits and errors for both cases. 

Afterward, we evaluate the secure key rate corresponding to the sifted bit and error sizes for the original PDTC, and the transformed PDTC. Results are shown in Figure \ref{fig:attachedPlotSigma}, and decoy parameters for each run are given in Table \ref{tab:asymmetricParams}. 

In Figure \ref{fig:attachedPlotSigma} we see the greatest impact when working at the strongest amounts of turbulence (indicated by increasing $\sigma$), and thus channel imbalance. In particular, for many of the plots shown, we see that dynamic attenuation gives the ability to generate a secure key when the atmospheric conditions would not otherwise allow it.
\begin{table}[ht!]
\centering
\resizebox{.5\textwidth}{!}{%
\begin{tabular}{||l|l|l|l|l|l|l|l||}
\hline
 $N$ & dB & $s$ & $\mu$ & $\nu$ & $P_{s}$ & $P_{\mu}$ & $P_{\nu}$\\
\hline
$10^{12}$ & 28 & 0.353 & 0.229 & 0.051 & 0.527 & 0.055 & 0.285\\
\hline
$10^{13}$ & 34 & 0.450 & 0.200 & 0.037 & 0.573 & 0.066 & 0.219\\
\hline
$10^{14}$ & 40 & 0.499 & 0.198 & 0.026 & 0.466 & 0.123 & 0.295\\
\hline
\end{tabular}%
}
\caption{Decoy parameters for each of our simulations.}
\label{tab:asymmetricParams}
\end{table}

Results show that dynamic attenuation gives a higher secure key rate when the turbulence induced transmittance fluctuation is more severe, as manifested by a higher $\sigma^2$. The largest benefit can be seen in shorter keys, though there is still an advantage in large ones.

\begin{figure}
    \centering
    \includegraphics[width=0.8\linewidth]{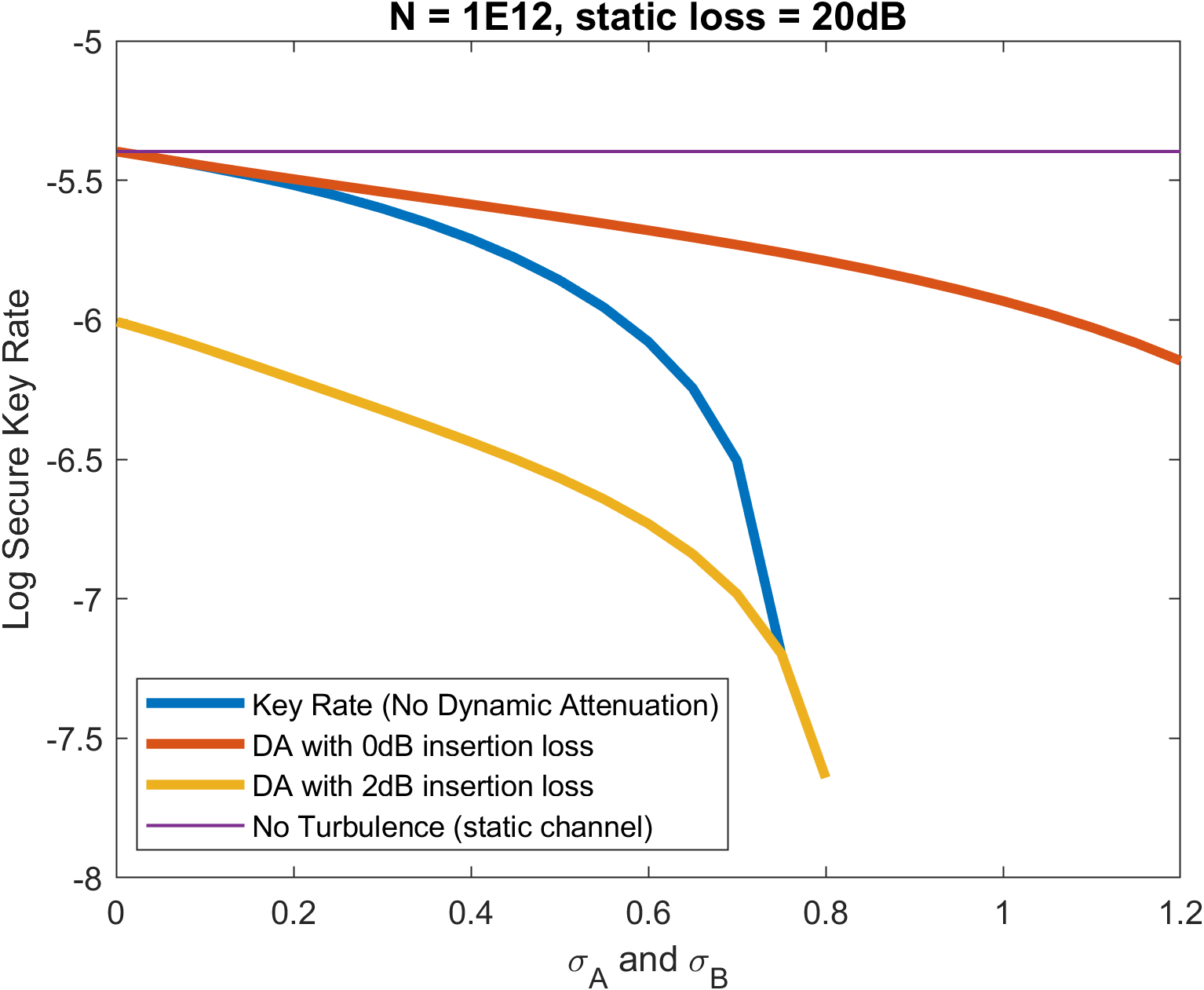}
    \caption{The impact of the minimum loss of VOA. Secret key rates with dynamic attenuation and minimum loss 0 dB (red) and 2 dB (yellow), and without dynamic attenuation (blue).}
    \label{fig:DA_insertionLoss}
\end{figure}

In the above simulations, we have assumed the minimum loss of the VOA can be set to 0 dB. However, if one considers to implement the high speed VOA using a commercial LiNbO3 amplitude modulator, the minimum insertion loss could be about 2-3 dB. There will be a penalty on secret key rate associated with the minimum insertion loss. Nevertheless, the advantage of the dynamic attenuation scheme at high turbulence remains, as shown in Figure \ref{fig:DA_insertionLoss}.

Figure \ref{fig:DA_insertionLoss} shows that dynamic attenuation with nonzero insertion loss is only beneficial beyond a certain level of turbulence. Thus, one should only use dynamic attenuation when the turbulence is high. At low data rates, this could be problematic, but for 1 GHz pulse rates \cite{Comandar2016}, it takes about 15-20 min to send $N = 10^{12}$. In this time frame, it is unlikely that turbulence will change drastically, as shown in \cite{Wang2015, Sprung2014, Qing2018}, so an entire experiment could be completed when turbulence is high and dynamic attenuation is most useful.

\section{Conclusion}\label{sec:con}

We introduced the seemingly paradoxical idea that one could enhance the secure key rate of an MDI QKD setup in turbulence by adding loss in one of the channels. We improved the HOM visibility by dynamically adding loss to balance the constantly fluctuating channel transmittances. This dynamic attenuation modified the original joint PDTC to one which was more favorable to MDI QKD.

It should be pointed out that in order to use fiber-based VOAs, one needs VOAs with extremely low loss, because much of the plot area in Figure \ref{fig:optimalAttenuation} shows an optimal attenuation of 0 dB whenever the channels' transmittances are close. Adding too much loss under these conditions could spoil advantages gained through dynamically attenuating less balanced channels. Nevertheless, as shown in Figure \ref{fig:DA_insertionLoss}, a strong advantage still remains for higher turbulence.

Our results show that secure keys can be obtained for much higher turbulence when one applies dynamic attenuation, especially when using short raw key lengths. We have shown that automated channel transmittance balancing is very helpful in extending MDI QKD's use in a highly turbulent environment. 

It would be interesting to derive the kernel $K$ used in our calculations (Eq.\ \eqref{eq:kernel}) rigorously, and apply our method in conjunction with a post selection process \cite{Cao2020}.

\acknowledgements

This material is based upon work supported by the U.S. Department of Energy, Office of Science, Office of Advanced Scientific Computing Research, through the Quantum Internet to Accelerate Scientific Discovery Program under Field Work Proposal 3ERKJ381.


\bibliographystyle{apsrev4-1}
\bibliography{dissertationBib}

\begin{thebibliography}{56}%
\makeatletter
\providecommand \@ifxundefined [1]{%
 \@ifx{#1\undefined}
}%
\providecommand \@ifnum [1]{%
 \ifnum #1\expandafter \@firstoftwo
 \else \expandafter \@secondoftwo
 \fi
}%
\providecommand \@ifx [1]{%
 \ifx #1\expandafter \@firstoftwo
 \else \expandafter \@secondoftwo
 \fi
}%
\providecommand \natexlab [1]{#1}%
\providecommand \enquote  [1]{``#1''}%
\providecommand \bibnamefont  [1]{#1}%
\providecommand \bibfnamefont [1]{#1}%
\providecommand \citenamefont [1]{#1}%
\providecommand \href@noop [0]{\@secondoftwo}%
\providecommand \href [0]{\begingroup \@sanitize@url \@href}%
\providecommand \@href[1]{\@@startlink{#1}\@@href}%
\providecommand \@@href[1]{\endgroup#1\@@endlink}%
\providecommand \@sanitize@url [0]{\catcode `\\12\catcode `\$12\catcode
  `\&12\catcode `\#12\catcode `\^12\catcode `\_12\catcode `\%12\relax}%
\providecommand \@@startlink[1]{}%
\providecommand \@@endlink[0]{}%
\providecommand \url  [0]{\begingroup\@sanitize@url \@url }%
\providecommand \@url [1]{\endgroup\@href {#1}{\urlprefix }}%
\providecommand \urlprefix  [0]{URL }%
\providecommand \Eprint [0]{\href }%
\providecommand \doibase [0]{http://dx.doi.org/}%
\providecommand \selectlanguage [0]{\@gobble}%
\providecommand \bibinfo  [0]{\@secondoftwo}%
\providecommand \bibfield  [0]{\@secondoftwo}%
\providecommand \translation [1]{[#1]}%
\providecommand \BibitemOpen [0]{}%
\providecommand \bibitemStop [0]{}%
\providecommand \bibitemNoStop [0]{.\EOS\space}%
\providecommand \EOS [0]{\spacefactor3000\relax}%
\providecommand \BibitemShut  [1]{\csname bibitem#1\endcsname}%
\let\auto@bib@innerbib\@empty
\bibitem [{\citenamefont {Bennett}\ and\ \citenamefont
  {Brassard}(1989)}]{Bennett1989}%
  \BibitemOpen
  \bibfield  {author} {\bibinfo {author} {\bibfnamefont {C.~H.}\ \bibnamefont
  {Bennett}}\ and\ \bibinfo {author} {\bibfnamefont {G.}~\bibnamefont
  {Brassard}},\ }\href {\doibase 10.1145/74074.74087} {\bibfield  {journal}
  {\bibinfo  {journal} {SIGACT News}\ }\textbf {\bibinfo {volume} {20}},\
  \bibinfo {pages} {78–80} (\bibinfo {year} {1989})}\BibitemShut {NoStop}%
\bibitem [{\citenamefont {Liao}\ \emph {et~al.}(2017)\citenamefont {Liao},
  \citenamefont {Cai}, \citenamefont {Liu}, \citenamefont {Zhang},
  \citenamefont {Li}, \citenamefont {Ren}, \citenamefont {Yin}, \citenamefont
  {Shen}, \citenamefont {Cao}, \citenamefont {Li}, \citenamefont {Li},
  \citenamefont {Chen}, \citenamefont {Sun}, \citenamefont {Jia}, \citenamefont
  {Wu}, \citenamefont {Jiang}, \citenamefont {Wang}, \citenamefont {Huang},
  \citenamefont {Wang}, \citenamefont {Zhou}, \citenamefont {Deng},
  \citenamefont {Xi}, \citenamefont {Ma}, \citenamefont {Hu}, \citenamefont
  {Zhang}, \citenamefont {Chen}, \citenamefont {Liu}, \citenamefont {Wang},
  \citenamefont {Zhu}, \citenamefont {Lu}, \citenamefont {Shu}, \citenamefont
  {Peng}, \citenamefont {Wang},\ and\ \citenamefont {Pan}}]{Liao2017a}%
  \BibitemOpen
  \bibfield  {author} {\bibinfo {author} {\bibfnamefont {S.-K.}\ \bibnamefont
  {Liao}}, \bibinfo {author} {\bibfnamefont {W.-Q.}\ \bibnamefont {Cai}},
  \bibinfo {author} {\bibfnamefont {W.-Y.}\ \bibnamefont {Liu}}, \bibinfo
  {author} {\bibfnamefont {L.}~\bibnamefont {Zhang}}, \bibinfo {author}
  {\bibfnamefont {Y.}~\bibnamefont {Li}}, \bibinfo {author} {\bibfnamefont
  {J.-G.}\ \bibnamefont {Ren}}, \bibinfo {author} {\bibfnamefont
  {J.}~\bibnamefont {Yin}}, \bibinfo {author} {\bibfnamefont {Q.}~\bibnamefont
  {Shen}}, \bibinfo {author} {\bibfnamefont {Y.}~\bibnamefont {Cao}}, \bibinfo
  {author} {\bibfnamefont {Z.-P.}\ \bibnamefont {Li}}, \bibinfo {author}
  {\bibfnamefont {F.-Z.}\ \bibnamefont {Li}}, \bibinfo {author} {\bibfnamefont
  {X.-W.}\ \bibnamefont {Chen}}, \bibinfo {author} {\bibfnamefont {L.-H.}\
  \bibnamefont {Sun}}, \bibinfo {author} {\bibfnamefont {J.-J.}\ \bibnamefont
  {Jia}}, \bibinfo {author} {\bibfnamefont {J.-C.}\ \bibnamefont {Wu}},
  \bibinfo {author} {\bibfnamefont {X.-J.}\ \bibnamefont {Jiang}}, \bibinfo
  {author} {\bibfnamefont {J.-F.}\ \bibnamefont {Wang}}, \bibinfo {author}
  {\bibfnamefont {Y.-M.}\ \bibnamefont {Huang}}, \bibinfo {author}
  {\bibfnamefont {Q.}~\bibnamefont {Wang}}, \bibinfo {author} {\bibfnamefont
  {Y.-L.}\ \bibnamefont {Zhou}}, \bibinfo {author} {\bibfnamefont
  {L.}~\bibnamefont {Deng}}, \bibinfo {author} {\bibfnamefont {T.}~\bibnamefont
  {Xi}}, \bibinfo {author} {\bibfnamefont {L.}~\bibnamefont {Ma}}, \bibinfo
  {author} {\bibfnamefont {T.}~\bibnamefont {Hu}}, \bibinfo {author}
  {\bibfnamefont {Q.}~\bibnamefont {Zhang}}, \bibinfo {author} {\bibfnamefont
  {Y.-A.}\ \bibnamefont {Chen}}, \bibinfo {author} {\bibfnamefont {N.-L.}\
  \bibnamefont {Liu}}, \bibinfo {author} {\bibfnamefont {X.-B.}\ \bibnamefont
  {Wang}}, \bibinfo {author} {\bibfnamefont {Z.-C.}\ \bibnamefont {Zhu}},
  \bibinfo {author} {\bibfnamefont {C.-Y.}\ \bibnamefont {Lu}}, \bibinfo
  {author} {\bibfnamefont {R.}~\bibnamefont {Shu}}, \bibinfo {author}
  {\bibfnamefont {C.-Z.}\ \bibnamefont {Peng}}, \bibinfo {author}
  {\bibfnamefont {J.-Y.}\ \bibnamefont {Wang}}, \ and\ \bibinfo {author}
  {\bibfnamefont {J.-W.}\ \bibnamefont {Pan}},\ }\href {\doibase
  10.1038/nature23655} {\bibfield  {journal} {\bibinfo  {journal} {Nature}\
  }\textbf {\bibinfo {volume} {549}},\ \bibinfo {pages} {43} (\bibinfo {year}
  {2017})}\BibitemShut {NoStop}%
\bibitem [{\citenamefont {Makarov}(2009)}]{Makarov2009}%
  \BibitemOpen
  \bibfield  {author} {\bibinfo {author} {\bibfnamefont {V.}~\bibnamefont
  {Makarov}},\ }\href@noop {} {\bibfield  {journal} {\bibinfo  {journal} {New
  Journal of Physics}\ }\textbf {\bibinfo {volume} {11}},\ \bibinfo {pages}
  {65003} (\bibinfo {year} {2009})}\BibitemShut {NoStop}%
\bibitem [{\citenamefont {Lydersen}\ \emph {et~al.}(2011)\citenamefont
  {Lydersen}, \citenamefont {Akhlaghi}, \citenamefont {Majedi}, \citenamefont
  {Skaar},\ and\ \citenamefont {Makarov}}]{Lydersen2011}%
  \BibitemOpen
  \bibfield  {author} {\bibinfo {author} {\bibfnamefont {L.}~\bibnamefont
  {Lydersen}}, \bibinfo {author} {\bibfnamefont {M.}~\bibnamefont {Akhlaghi}},
  \bibinfo {author} {\bibfnamefont {A.}~\bibnamefont {Majedi}}, \bibinfo
  {author} {\bibfnamefont {J.}~\bibnamefont {Skaar}}, \ and\ \bibinfo {author}
  {\bibfnamefont {V.}~\bibnamefont {Makarov}},\ }\href@noop {} {\bibfield
  {journal} {\bibinfo  {journal} {New Journal of Physics}\ }\textbf {\bibinfo
  {volume} {13}},\ \bibinfo {pages} {113042} (\bibinfo {year}
  {2011})}\BibitemShut {NoStop}%
\bibitem [{\citenamefont {Qi}\ \emph {et~al.}(2005)\citenamefont {Qi},
  \citenamefont {Fung}, \citenamefont {Lo},\ and\ \citenamefont
  {Ma}}]{qi2005time}%
  \BibitemOpen
  \bibfield  {author} {\bibinfo {author} {\bibfnamefont {B.}~\bibnamefont
  {Qi}}, \bibinfo {author} {\bibfnamefont {C.-H.~F.}\ \bibnamefont {Fung}},
  \bibinfo {author} {\bibfnamefont {H.-K.}\ \bibnamefont {Lo}}, \ and\ \bibinfo
  {author} {\bibfnamefont {X.}~\bibnamefont {Ma}},\ }\href@noop {} {\bibfield
  {journal} {\bibinfo  {journal} {arXiv preprint quant-ph/0512080}\ } (\bibinfo
  {year} {2005})}\BibitemShut {NoStop}%
\bibitem [{\citenamefont {Pinheiro}\ \emph {et~al.}(2018)\citenamefont
  {Pinheiro}, \citenamefont {Pereira}, \citenamefont {Chaiwongkhot},
  \citenamefont {Sajeed}, \citenamefont {Horn}, \citenamefont {Bourgoin},
  \citenamefont {Jennewein}, \citenamefont {Lütkenhaus},\ and\ \citenamefont
  {Makarov}}]{Pinheiro2018}%
  \BibitemOpen
  \bibfield  {author} {\bibinfo {author} {\bibfnamefont {P.}~\bibnamefont
  {Pinheiro}}, \bibinfo {author} {\bibfnamefont {P.}~\bibnamefont {Pereira}},
  \bibinfo {author} {\bibfnamefont {S.}~\bibnamefont {Chaiwongkhot}}, \bibinfo
  {author} {\bibfnamefont {R.}~\bibnamefont {Sajeed}}, \bibinfo {author}
  {\bibfnamefont {J.-P.}\ \bibnamefont {Horn}}, \bibinfo {author}
  {\bibfnamefont {T.}~\bibnamefont {Bourgoin}}, \bibinfo {author}
  {\bibfnamefont {N.}~\bibnamefont {Jennewein}}, \bibinfo {author}
  {\bibfnamefont {V.}~\bibnamefont {Lütkenhaus}}, \ and\ \bibinfo {author}
  {\bibnamefont {Makarov}},\ }\href@noop {} {\bibfield  {journal} {\bibinfo
  {journal} {Optics express}\ }\textbf {\bibinfo {volume} {26}},\ \bibinfo
  {pages} {21020} (\bibinfo {year} {2018})}\BibitemShut {NoStop}%
\bibitem [{\citenamefont {Jain}\ \emph {et~al.}(2016)\citenamefont {Jain},
  \citenamefont {Stiller}, \citenamefont {Khan}, \citenamefont {Elser},
  \citenamefont {Marquardt},\ and\ \citenamefont {Leuchs}}]{Jain2016}%
  \BibitemOpen
  \bibfield  {author} {\bibinfo {author} {\bibfnamefont {N.}~\bibnamefont
  {Jain}}, \bibinfo {author} {\bibfnamefont {B.}~\bibnamefont {Stiller}},
  \bibinfo {author} {\bibfnamefont {I.}~\bibnamefont {Khan}}, \bibinfo {author}
  {\bibfnamefont {D.}~\bibnamefont {Elser}}, \bibinfo {author} {\bibfnamefont
  {C.}~\bibnamefont {Marquardt}}, \ and\ \bibinfo {author} {\bibfnamefont
  {G.}~\bibnamefont {Leuchs}},\ }\href {\doibase 10.1080/00107514.2016.1148333}
  {\bibfield  {journal} {\bibinfo  {journal} {arXiv:1512.07990}\ } (\bibinfo
  {year} {2016}),\ 10.1080/00107514.2016.1148333}\BibitemShut {NoStop}%
\bibitem [{\citenamefont {Ac{\'\i}n}\ \emph {et~al.}(2007)\citenamefont
  {Ac{\'\i}n}, \citenamefont {Brunner}, \citenamefont {Gisin}, \citenamefont
  {Massar}, \citenamefont {Pironio},\ and\ \citenamefont
  {Scarani}}]{acin2007device}%
  \BibitemOpen
  \bibfield  {author} {\bibinfo {author} {\bibfnamefont {A.}~\bibnamefont
  {Ac{\'\i}n}}, \bibinfo {author} {\bibfnamefont {N.}~\bibnamefont {Brunner}},
  \bibinfo {author} {\bibfnamefont {N.}~\bibnamefont {Gisin}}, \bibinfo
  {author} {\bibfnamefont {S.}~\bibnamefont {Massar}}, \bibinfo {author}
  {\bibfnamefont {S.}~\bibnamefont {Pironio}}, \ and\ \bibinfo {author}
  {\bibfnamefont {V.}~\bibnamefont {Scarani}},\ }\href@noop {} {\bibfield
  {journal} {\bibinfo  {journal} {Physical Review Letters}\ }\textbf {\bibinfo
  {volume} {98}},\ \bibinfo {pages} {230501} (\bibinfo {year}
  {2007})}\BibitemShut {NoStop}%
\bibitem [{\citenamefont {Barrett}\ \emph {et~al.}(2005)\citenamefont
  {Barrett}, \citenamefont {Hardy},\ and\ \citenamefont
  {Kent}}]{barrett2005no}%
  \BibitemOpen
  \bibfield  {author} {\bibinfo {author} {\bibfnamefont {J.}~\bibnamefont
  {Barrett}}, \bibinfo {author} {\bibfnamefont {L.}~\bibnamefont {Hardy}}, \
  and\ \bibinfo {author} {\bibfnamefont {A.}~\bibnamefont {Kent}},\ }\href@noop
  {} {\bibfield  {journal} {\bibinfo  {journal} {Phys. Rev. Lett.}\ }\textbf
  {\bibinfo {volume} {95}},\ \bibinfo {pages} {010503} (\bibinfo {year}
  {2005})}\BibitemShut {NoStop}%
\bibitem [{\citenamefont {Mayers}\ and\ \citenamefont
  {Yao}(1998)}]{mayers1998quantum}%
  \BibitemOpen
  \bibfield  {author} {\bibinfo {author} {\bibfnamefont {D.}~\bibnamefont
  {Mayers}}\ and\ \bibinfo {author} {\bibfnamefont {A.}~\bibnamefont {Yao}},\
  }in\ \href@noop {} {\emph {\bibinfo {booktitle} {Proceedings 39th Annual
  Symposium on Foundations of Computer Science (Cat. No. 98CB36280)}}}\
  (\bibinfo {organization} {IEEE},\ \bibinfo {year} {1998})\ pp.\ \bibinfo
  {pages} {503--509}\BibitemShut {NoStop}%
\bibitem [{\citenamefont {Bell}(1964)}]{Bell1964}%
  \BibitemOpen
  \bibfield  {author} {\bibinfo {author} {\bibfnamefont {J.~S.}\ \bibnamefont
  {Bell}},\ }\href@noop {} {\bibfield  {journal} {\bibinfo  {journal} {Physics
  Physique Fizika}\ }\textbf {\bibinfo {volume} {1}},\ \bibinfo {pages} {195}
  (\bibinfo {year} {1964})}\BibitemShut {NoStop}%
\bibitem [{\citenamefont {Garg}\ and\ \citenamefont {Mermin}(1987)}]{Garg}%
  \BibitemOpen
  \bibfield  {author} {\bibinfo {author} {\bibfnamefont {A.}~\bibnamefont
  {Garg}}\ and\ \bibinfo {author} {\bibfnamefont {N.~D.}\ \bibnamefont
  {Mermin}},\ }\href {\doibase 10.1103/physrevd.35.3831} {\bibfield  {journal}
  {\bibinfo  {journal} {Phys. Rev. D}\ }\textbf {\bibinfo {volume} {35}},\
  \bibinfo {pages} {3831} (\bibinfo {year} {1987})}\BibitemShut {NoStop}%
\bibitem [{\citenamefont {Zhang}\ \emph {et~al.}(2021)\citenamefont {Zhang},
  \citenamefont {van Leent}, \citenamefont {Redeker}, \citenamefont {Garthoff},
  \citenamefont {Schwonnek}, \citenamefont {Fertig}, \citenamefont {Eppelt},
  \citenamefont {Scarani}, \citenamefont {Lim},\ and\ \citenamefont
  {Weinfurter}}]{zhang2021experimental}%
  \BibitemOpen
  \bibfield  {author} {\bibinfo {author} {\bibfnamefont {W.}~\bibnamefont
  {Zhang}}, \bibinfo {author} {\bibfnamefont {T.}~\bibnamefont {van Leent}},
  \bibinfo {author} {\bibfnamefont {K.}~\bibnamefont {Redeker}}, \bibinfo
  {author} {\bibfnamefont {R.}~\bibnamefont {Garthoff}}, \bibinfo {author}
  {\bibfnamefont {R.}~\bibnamefont {Schwonnek}}, \bibinfo {author}
  {\bibfnamefont {F.}~\bibnamefont {Fertig}}, \bibinfo {author} {\bibfnamefont
  {S.}~\bibnamefont {Eppelt}}, \bibinfo {author} {\bibfnamefont
  {V.}~\bibnamefont {Scarani}}, \bibinfo {author} {\bibfnamefont {C.~C.-W.}\
  \bibnamefont {Lim}}, \ and\ \bibinfo {author} {\bibfnamefont
  {H.}~\bibnamefont {Weinfurter}},\ }\href@noop {} {\bibfield  {journal}
  {\bibinfo  {journal} {arXiv preprint arXiv:2110.00575}\ } (\bibinfo {year}
  {2021})}\BibitemShut {NoStop}%
\bibitem [{\citenamefont {Lo}\ \emph {et~al.}(2012)\citenamefont {Lo},
  \citenamefont {Curty},\ and\ \citenamefont {Qi}}]{Lo2012}%
  \BibitemOpen
  \bibfield  {author} {\bibinfo {author} {\bibfnamefont {H.-K.}\ \bibnamefont
  {Lo}}, \bibinfo {author} {\bibfnamefont {M.}~\bibnamefont {Curty}}, \ and\
  \bibinfo {author} {\bibfnamefont {B.}~\bibnamefont {Qi}},\ }\href {\doibase
  10.1103/PhysRevLett.108.130503} {\bibfield  {journal} {\bibinfo  {journal}
  {Phys. Rev. Lett.}\ }\textbf {\bibinfo {volume} {108}},\ \bibinfo {pages}
  {130503} (\bibinfo {year} {2012})}\BibitemShut {NoStop}%
\bibitem [{\citenamefont {Rubenok}\ \emph {et~al.}(2013)\citenamefont
  {Rubenok}, \citenamefont {Slater}, \citenamefont {Chan}, \citenamefont
  {Lucio-Martinez},\ and\ \citenamefont {Tittel}}]{Rubenok2013}%
  \BibitemOpen
  \bibfield  {author} {\bibinfo {author} {\bibfnamefont {A.}~\bibnamefont
  {Rubenok}}, \bibinfo {author} {\bibfnamefont {J.~A.}\ \bibnamefont {Slater}},
  \bibinfo {author} {\bibfnamefont {P.}~\bibnamefont {Chan}}, \bibinfo {author}
  {\bibfnamefont {I.}~\bibnamefont {Lucio-Martinez}}, \ and\ \bibinfo {author}
  {\bibfnamefont {W.}~\bibnamefont {Tittel}},\ }\href {\doibase
  10.1103/PhysRevLett.111.130501} {\bibfield  {journal} {\bibinfo  {journal}
  {Physical review letters}\ }\textbf {\bibinfo {volume} {111}},\ \bibinfo
  {pages} {130501} (\bibinfo {year} {2013})}\BibitemShut {NoStop}%
\bibitem [{\citenamefont {Tang}\ \emph
  {et~al.}(2014{\natexlab{a}})\citenamefont {Tang}, \citenamefont {Liao},
  \citenamefont {Xu}, \citenamefont {Qi}, \citenamefont {Qian},\ and\
  \citenamefont {Lo}}]{tang2014experimental}%
  \BibitemOpen
  \bibfield  {author} {\bibinfo {author} {\bibfnamefont {Z.}~\bibnamefont
  {Tang}}, \bibinfo {author} {\bibfnamefont {Z.}~\bibnamefont {Liao}}, \bibinfo
  {author} {\bibfnamefont {F.}~\bibnamefont {Xu}}, \bibinfo {author}
  {\bibfnamefont {B.}~\bibnamefont {Qi}}, \bibinfo {author} {\bibfnamefont
  {L.}~\bibnamefont {Qian}}, \ and\ \bibinfo {author} {\bibfnamefont {H.-K.}\
  \bibnamefont {Lo}},\ }\href@noop {} {\bibfield  {journal} {\bibinfo
  {journal} {Physical review letters}\ }\textbf {\bibinfo {volume} {112}},\
  \bibinfo {pages} {190503} (\bibinfo {year} {2014}{\natexlab{a}})}\BibitemShut
  {NoStop}%
\bibitem [{\citenamefont {Yin}\ \emph {et~al.}(2016)\citenamefont {Yin},
  \citenamefont {Chen}, \citenamefont {Yu}, \citenamefont {Liu}, \citenamefont
  {You}, \citenamefont {Zhou}, \citenamefont {Chen}, \citenamefont {Mao},
  \citenamefont {Huang}, \citenamefont {Zhang} \emph
  {et~al.}}]{yin2016measurement}%
  \BibitemOpen
  \bibfield  {author} {\bibinfo {author} {\bibfnamefont {H.-L.}\ \bibnamefont
  {Yin}}, \bibinfo {author} {\bibfnamefont {T.-Y.}\ \bibnamefont {Chen}},
  \bibinfo {author} {\bibfnamefont {Z.-W.}\ \bibnamefont {Yu}}, \bibinfo
  {author} {\bibfnamefont {H.}~\bibnamefont {Liu}}, \bibinfo {author}
  {\bibfnamefont {L.-X.}\ \bibnamefont {You}}, \bibinfo {author} {\bibfnamefont
  {Y.-H.}\ \bibnamefont {Zhou}}, \bibinfo {author} {\bibfnamefont {S.-J.}\
  \bibnamefont {Chen}}, \bibinfo {author} {\bibfnamefont {Y.}~\bibnamefont
  {Mao}}, \bibinfo {author} {\bibfnamefont {M.-Q.}\ \bibnamefont {Huang}},
  \bibinfo {author} {\bibfnamefont {W.-J.}\ \bibnamefont {Zhang}},  \emph
  {et~al.},\ }\href@noop {} {\bibfield  {journal} {\bibinfo  {journal}
  {Physical review letters}\ }\textbf {\bibinfo {volume} {117}},\ \bibinfo
  {pages} {190501} (\bibinfo {year} {2016})}\BibitemShut {NoStop}%
\bibitem [{\citenamefont {Wei}\ \emph {et~al.}(2020)\citenamefont {Wei},
  \citenamefont {Li}, \citenamefont {Tan}, \citenamefont {Li}, \citenamefont
  {Min}, \citenamefont {Zhang}, \citenamefont {Li}, \citenamefont {You},
  \citenamefont {Wang}, \citenamefont {Jiang}, \citenamefont {Chen},
  \citenamefont {Liao}, \citenamefont {Peng}, \citenamefont {Xu},\ and\
  \citenamefont {Pan}}]{Wei2020}%
  \BibitemOpen
  \bibfield  {author} {\bibinfo {author} {\bibfnamefont {K.}~\bibnamefont
  {Wei}}, \bibinfo {author} {\bibfnamefont {W.}~\bibnamefont {Li}}, \bibinfo
  {author} {\bibfnamefont {H.}~\bibnamefont {Tan}}, \bibinfo {author}
  {\bibfnamefont {Y.}~\bibnamefont {Li}}, \bibinfo {author} {\bibfnamefont
  {H.}~\bibnamefont {Min}}, \bibinfo {author} {\bibfnamefont {W.-J.}\
  \bibnamefont {Zhang}}, \bibinfo {author} {\bibfnamefont {H.}~\bibnamefont
  {Li}}, \bibinfo {author} {\bibfnamefont {L.}~\bibnamefont {You}}, \bibinfo
  {author} {\bibfnamefont {Z.}~\bibnamefont {Wang}}, \bibinfo {author}
  {\bibfnamefont {X.}~\bibnamefont {Jiang}}, \bibinfo {author} {\bibfnamefont
  {T.-Y.}\ \bibnamefont {Chen}}, \bibinfo {author} {\bibfnamefont {S.-K.}\
  \bibnamefont {Liao}}, \bibinfo {author} {\bibfnamefont {C.-Z.}\ \bibnamefont
  {Peng}}, \bibinfo {author} {\bibfnamefont {F.}~\bibnamefont {Xu}}, \ and\
  \bibinfo {author} {\bibfnamefont {J.-W.}\ \bibnamefont {Pan}},\ }\href
  {\doibase 10.1103/PhysRevX.10.031030} {\bibfield  {journal} {\bibinfo
  {journal} {Physical Review X}\ }\textbf {\bibinfo {volume} {10}},\ \bibinfo
  {eid} {031030} (\bibinfo {year} {2020})},\ \Eprint
  {http://arxiv.org/abs/1911.00690} {arXiv:1911.00690 [quant-ph]} \BibitemShut
  {NoStop}%
\bibitem [{\citenamefont {Valivarthi}\ \emph {et~al.}(2017)\citenamefont
  {Valivarthi}, \citenamefont {Zhou}, \citenamefont {John}, \citenamefont
  {Marsili}, \citenamefont {Verma}, \citenamefont {Shaw}, \citenamefont {Nam},
  \citenamefont {Oblak},\ and\ \citenamefont {Tittel}}]{valivarthi2017cost}%
  \BibitemOpen
  \bibfield  {author} {\bibinfo {author} {\bibfnamefont {R.}~\bibnamefont
  {Valivarthi}}, \bibinfo {author} {\bibfnamefont {Q.}~\bibnamefont {Zhou}},
  \bibinfo {author} {\bibfnamefont {C.}~\bibnamefont {John}}, \bibinfo {author}
  {\bibfnamefont {F.}~\bibnamefont {Marsili}}, \bibinfo {author} {\bibfnamefont
  {V.~B.}\ \bibnamefont {Verma}}, \bibinfo {author} {\bibfnamefont {M.~D.}\
  \bibnamefont {Shaw}}, \bibinfo {author} {\bibfnamefont {S.~W.}\ \bibnamefont
  {Nam}}, \bibinfo {author} {\bibfnamefont {D.}~\bibnamefont {Oblak}}, \ and\
  \bibinfo {author} {\bibfnamefont {W.}~\bibnamefont {Tittel}},\ }\href@noop {}
  {\bibfield  {journal} {\bibinfo  {journal} {Quantum Science and Technology}\
  }\textbf {\bibinfo {volume} {2}},\ \bibinfo {pages} {04LT01} (\bibinfo {year}
  {2017})}\BibitemShut {NoStop}%
\bibitem [{\citenamefont {Hong}\ \emph {et~al.}(1987)\citenamefont {Hong},
  \citenamefont {Ou},\ and\ \citenamefont {Mandel}}]{Hong1987}%
  \BibitemOpen
  \bibfield  {author} {\bibinfo {author} {\bibfnamefont {C.-K.}\ \bibnamefont
  {Hong}}, \bibinfo {author} {\bibfnamefont {Z.-Y.}\ \bibnamefont {Ou}}, \ and\
  \bibinfo {author} {\bibfnamefont {L.}~\bibnamefont {Mandel}},\ }\href@noop {}
  {\bibfield  {journal} {\bibinfo  {journal} {Physical review letters}\
  }\textbf {\bibinfo {volume} {59}},\ \bibinfo {pages} {2044} (\bibinfo {year}
  {1987})}\BibitemShut {NoStop}%
\bibitem [{\citenamefont {{Wang}}\ \emph {et~al.}(2017)\citenamefont {{Wang}},
  \citenamefont {{Wang}}, \citenamefont {{Chen}}, \citenamefont {{Wang}},
  \citenamefont {{Chen}}, \citenamefont {{Yin}}, \citenamefont {{He}},
  \citenamefont {{Guo}},\ and\ \citenamefont {{Han}}}]{Wang2017}%
  \BibitemOpen
  \bibfield  {author} {\bibinfo {author} {\bibfnamefont {C.}~\bibnamefont
  {{Wang}}}, \bibinfo {author} {\bibfnamefont {F.}~\bibnamefont {{Wang}}},
  \bibinfo {author} {\bibfnamefont {H.}~\bibnamefont {{Chen}}}, \bibinfo
  {author} {\bibfnamefont {S.}~\bibnamefont {{Wang}}}, \bibinfo {author}
  {\bibfnamefont {W.}~\bibnamefont {{Chen}}}, \bibinfo {author} {\bibfnamefont
  {Z.}~\bibnamefont {{Yin}}}, \bibinfo {author} {\bibfnamefont
  {D.}~\bibnamefont {{He}}}, \bibinfo {author} {\bibfnamefont {G.}~\bibnamefont
  {{Guo}}}, \ and\ \bibinfo {author} {\bibfnamefont {Z.}~\bibnamefont
  {{Han}}},\ }\href {\doibase 10.1109/JLT.2017.2764140} {\bibfield  {journal}
  {\bibinfo  {journal} {Journal of Lightwave Technology}\ }\textbf {\bibinfo
  {volume} {35}},\ \bibinfo {pages} {4996} (\bibinfo {year}
  {2017})}\BibitemShut {NoStop}%
\bibitem [{\citenamefont {{Moschandreou}}\ \emph {et~al.}(2018)\citenamefont
  {{Moschandreou}}, \citenamefont {{Garcia}}, \citenamefont {{Rollick}},
  \citenamefont {{Qi}}, \citenamefont {{Pooser}},\ and\ \citenamefont
  {{Siopsis}}}]{Moschandreou2018}%
  \BibitemOpen
  \bibfield  {author} {\bibinfo {author} {\bibfnamefont {E.}~\bibnamefont
  {{Moschandreou}}}, \bibinfo {author} {\bibfnamefont {J.~I.}\ \bibnamefont
  {{Garcia}}}, \bibinfo {author} {\bibfnamefont {B.~J.}\ \bibnamefont
  {{Rollick}}}, \bibinfo {author} {\bibfnamefont {B.}~\bibnamefont {{Qi}}},
  \bibinfo {author} {\bibfnamefont {R.}~\bibnamefont {{Pooser}}}, \ and\
  \bibinfo {author} {\bibfnamefont {G.}~\bibnamefont {{Siopsis}}},\ }\href
  {\doibase 10.1109/JLT.2018.2850282} {\bibfield  {journal} {\bibinfo
  {journal} {Journal of Lightwave Technology}\ }\textbf {\bibinfo {volume}
  {36}},\ \bibinfo {pages} {3752} (\bibinfo {year} {2018})}\BibitemShut
  {NoStop}%
\bibitem [{\citenamefont {Xu}\ \emph {et~al.}(2013)\citenamefont {Xu},
  \citenamefont {Curty}, \citenamefont {Qi},\ and\ \citenamefont
  {Lo}}]{Xu2013}%
  \BibitemOpen
  \bibfield  {author} {\bibinfo {author} {\bibfnamefont {F.}~\bibnamefont
  {Xu}}, \bibinfo {author} {\bibfnamefont {M.}~\bibnamefont {Curty}}, \bibinfo
  {author} {\bibfnamefont {B.}~\bibnamefont {Qi}}, \ and\ \bibinfo {author}
  {\bibfnamefont {H.-K.}\ \bibnamefont {Lo}},\ }\href {\doibase
  10.1088/1367-2630/15/11/113007} {\bibfield  {journal} {\bibinfo  {journal}
  {New Journal of Physics}\ }\textbf {\bibinfo {volume} {15}},\ \bibinfo
  {pages} {113007} (\bibinfo {year} {2013})}\BibitemShut {NoStop}%
\bibitem [{\citenamefont {Wang}\ \emph
  {et~al.}(2019{\natexlab{a}})\citenamefont {Wang}, \citenamefont {Xu},\ and\
  \citenamefont {Lo}}]{Wang2019a}%
  \BibitemOpen
  \bibfield  {author} {\bibinfo {author} {\bibfnamefont {W.}~\bibnamefont
  {Wang}}, \bibinfo {author} {\bibfnamefont {F.}~\bibnamefont {Xu}}, \ and\
  \bibinfo {author} {\bibfnamefont {H.-K.}\ \bibnamefont {Lo}},\ }\href@noop {}
  {\bibfield  {journal} {\bibinfo  {journal} {Physical Review X}\ }\textbf
  {\bibinfo {volume} {9}},\ \bibinfo {pages} {41012} (\bibinfo {year}
  {2019}{\natexlab{a}})}\BibitemShut {NoStop}%
\bibitem [{\citenamefont {Osche}(2002)}]{Osche2002}%
  \BibitemOpen
  \bibfield  {author} {\bibinfo {author} {\bibfnamefont {G.~R.}\ \bibnamefont
  {Osche}},\ }\href@noop {} {\emph {\bibinfo {title} {Optical detection theory
  for laser applications}}}\ (\bibinfo  {publisher} {John Wiley and Sons
  Inc.},\ \bibinfo {year} {2002})\BibitemShut {NoStop}%
\bibitem [{\citenamefont {Erven}\ \emph {et~al.}(2012)\citenamefont {Erven},
  \citenamefont {Heim}, \citenamefont {Meyer-Scott}, \citenamefont {Bourgoin},
  \citenamefont {Laflamme}, \citenamefont {Weihs},\ and\ \citenamefont
  {Jennewein}}]{Erven_2012}%
  \BibitemOpen
  \bibfield  {author} {\bibinfo {author} {\bibfnamefont {C.}~\bibnamefont
  {Erven}}, \bibinfo {author} {\bibfnamefont {B.}~\bibnamefont {Heim}},
  \bibinfo {author} {\bibfnamefont {E.}~\bibnamefont {Meyer-Scott}}, \bibinfo
  {author} {\bibfnamefont {J.~P.}\ \bibnamefont {Bourgoin}}, \bibinfo {author}
  {\bibfnamefont {R.}~\bibnamefont {Laflamme}}, \bibinfo {author}
  {\bibfnamefont {G.}~\bibnamefont {Weihs}}, \ and\ \bibinfo {author}
  {\bibfnamefont {T.}~\bibnamefont {Jennewein}},\ }\href {\doibase
  10.1088/1367-2630/14/12/123018} {\bibfield  {journal} {\bibinfo  {journal}
  {New Journal of Physics}\ }\textbf {\bibinfo {volume} {14}},\ \bibinfo
  {pages} {123018} (\bibinfo {year} {2012})}\BibitemShut {NoStop}%
\bibitem [{\citenamefont {Capraro}\ \emph {et~al.}(2012)\citenamefont
  {Capraro}, \citenamefont {Tomaello}, \citenamefont {Dall'Arche},
  \citenamefont {Gerlin}, \citenamefont {Ursin}, \citenamefont {Vallone},\ and\
  \citenamefont {Villoresi}}]{Capraro2012}%
  \BibitemOpen
  \bibfield  {author} {\bibinfo {author} {\bibfnamefont {I.}~\bibnamefont
  {Capraro}}, \bibinfo {author} {\bibfnamefont {A.}~\bibnamefont {Tomaello}},
  \bibinfo {author} {\bibfnamefont {A.}~\bibnamefont {Dall'Arche}}, \bibinfo
  {author} {\bibfnamefont {F.}~\bibnamefont {Gerlin}}, \bibinfo {author}
  {\bibfnamefont {R.}~\bibnamefont {Ursin}}, \bibinfo {author} {\bibfnamefont
  {G.}~\bibnamefont {Vallone}}, \ and\ \bibinfo {author} {\bibfnamefont
  {P.}~\bibnamefont {Villoresi}},\ }\href {\doibase
  10.1103/PhysRevLett.109.200502} {\bibfield  {journal} {\bibinfo  {journal}
  {Phys. Rev. Lett.}\ }\textbf {\bibinfo {volume} {109}} (\bibinfo {year}
  {2012}),\ 10.1103/PhysRevLett.109.200502},\ \Eprint
  {http://arxiv.org/abs/http://arxiv.org/abs/1207.6931v1}
  {http://arxiv.org/abs/1207.6931v1} \BibitemShut {NoStop}%
\bibitem [{\citenamefont {Vallone}\ \emph {et~al.}(2015)\citenamefont
  {Vallone}, \citenamefont {Marangon}, \citenamefont {Canale}, \citenamefont
  {Savorgnan}, \citenamefont {Bacco}, \citenamefont {Barbieri}, \citenamefont
  {Calimani}, \citenamefont {Barbieri}, \citenamefont {Laurenti},\ and\
  \citenamefont {Villoresi}}]{Vallone2015}%
  \BibitemOpen
  \bibfield  {author} {\bibinfo {author} {\bibfnamefont {G.}~\bibnamefont
  {Vallone}}, \bibinfo {author} {\bibfnamefont {D.~G.}\ \bibnamefont
  {Marangon}}, \bibinfo {author} {\bibfnamefont {M.}~\bibnamefont {Canale}},
  \bibinfo {author} {\bibfnamefont {I.}~\bibnamefont {Savorgnan}}, \bibinfo
  {author} {\bibfnamefont {D.}~\bibnamefont {Bacco}}, \bibinfo {author}
  {\bibfnamefont {M.}~\bibnamefont {Barbieri}}, \bibinfo {author}
  {\bibfnamefont {S.}~\bibnamefont {Calimani}}, \bibinfo {author}
  {\bibfnamefont {C.}~\bibnamefont {Barbieri}}, \bibinfo {author}
  {\bibfnamefont {N.}~\bibnamefont {Laurenti}}, \ and\ \bibinfo {author}
  {\bibfnamefont {P.}~\bibnamefont {Villoresi}},\ }\href {\doibase
  10.1103/PhysRevA.91.042320} {\bibfield  {journal} {\bibinfo  {journal} {Phys.
  Rev. A}\ }\textbf {\bibinfo {volume} {91}},\ \bibinfo {pages} {042320}
  (\bibinfo {year} {2015})}\BibitemShut {NoStop}%
\bibitem [{\citenamefont {Wang}\ \emph {et~al.}(2018)\citenamefont {Wang},
  \citenamefont {Xu},\ and\ \citenamefont {Lo}}]{Wang2018}%
  \BibitemOpen
  \bibfield  {author} {\bibinfo {author} {\bibfnamefont {W.}~\bibnamefont
  {Wang}}, \bibinfo {author} {\bibfnamefont {F.}~\bibnamefont {Xu}}, \ and\
  \bibinfo {author} {\bibfnamefont {H.-K.}\ \bibnamefont {Lo}},\ }\href
  {\doibase 10.1103/PhysRevA.97.032337} {\bibfield  {journal} {\bibinfo
  {journal} {Phys. Rev. A}\ }\textbf {\bibinfo {volume} {97}},\ \bibinfo
  {pages} {032337} (\bibinfo {year} {2018})}\BibitemShut {NoStop}%
\bibitem [{\citenamefont {Moschandreou}\ \emph {et~al.}(2021)\citenamefont
  {Moschandreou}, \citenamefont {Rollick}, \citenamefont {Qi},\ and\
  \citenamefont {Siopsis}}]{Moschandreou2021}%
  \BibitemOpen
  \bibfield  {author} {\bibinfo {author} {\bibfnamefont {E.}~\bibnamefont
  {Moschandreou}}, \bibinfo {author} {\bibfnamefont {B.~J.}\ \bibnamefont
  {Rollick}}, \bibinfo {author} {\bibfnamefont {B.}~\bibnamefont {Qi}}, \ and\
  \bibinfo {author} {\bibfnamefont {G.}~\bibnamefont {Siopsis}},\ }\href
  {\doibase 10.1103/physreva.103.032614} {\bibfield  {journal} {\bibinfo
  {journal} {Phys. Rev. A}\ }\textbf {\bibinfo {volume} {103}},\ \bibinfo
  {pages} {032614} (\bibinfo {year} {2021})}\BibitemShut {NoStop}%
\bibitem [{\citenamefont {Cao}(2020)}]{Cao2020}%
  \BibitemOpen
  \bibfield  {author} {\bibinfo {author} {\bibfnamefont {Y.}~\bibnamefont
  {Cao}},\ }\href@noop {} {\bibfield  {journal} {\bibinfo  {journal} {Physical
  Review Letters}\ }\textbf {\bibinfo {volume} {125}},\ \bibinfo {pages}
  {260503} (\bibinfo {year} {2020})}\BibitemShut {NoStop}%
\bibitem [{\citenamefont {Wang}\ \emph
  {et~al.}(2019{\natexlab{b}})\citenamefont {Wang}, \citenamefont {Xu},\ and\
  \citenamefont {Lo}}]{Wang2019}%
  \BibitemOpen
  \bibfield  {author} {\bibinfo {author} {\bibfnamefont {W.}~\bibnamefont
  {Wang}}, \bibinfo {author} {\bibfnamefont {F.}~\bibnamefont {Xu}}, \ and\
  \bibinfo {author} {\bibfnamefont {H.-K.}\ \bibnamefont {Lo}},\ }\href@noop {}
  {\bibfield  {journal} {\bibinfo  {journal} {arXiv preprint arXiv:1910.10137}\
  } (\bibinfo {year} {2019}{\natexlab{b}})},\ \Eprint
  {http://arxiv.org/abs/arXiv:1910.10137} {arXiv:1910.10137} \BibitemShut
  {NoStop}%
\bibitem [{\citenamefont {Zhu}\ \emph {et~al.}(2018)\citenamefont {Zhu},
  \citenamefont {Chen}, \citenamefont {Zhao}, \citenamefont {Zhang},\ and\
  \citenamefont {Xi}}]{Zhu2018}%
  \BibitemOpen
  \bibfield  {author} {\bibinfo {author} {\bibfnamefont {Z.-D.}\ \bibnamefont
  {Zhu}}, \bibinfo {author} {\bibfnamefont {D.}~\bibnamefont {Chen}}, \bibinfo
  {author} {\bibfnamefont {S.-H.}\ \bibnamefont {Zhao}}, \bibinfo {author}
  {\bibfnamefont {Q.-H.}\ \bibnamefont {Zhang}}, \ and\ \bibinfo {author}
  {\bibfnamefont {J.-H.}\ \bibnamefont {Xi}},\ }\href {\doibase
  10.1007/s11128-018-2146-9} {\bibfield  {journal} {\bibinfo  {journal}
  {Quantum Information Processing}\ }\textbf {\bibinfo {volume} {18}},\
  \bibinfo {pages} {33} (\bibinfo {year} {2018})}\BibitemShut {NoStop}%
\bibitem [{\citenamefont {Biham}\ \emph {et~al.}(1996)\citenamefont {Biham},
  \citenamefont {Huttner},\ and\ \citenamefont {Mor}}]{biham1996}%
  \BibitemOpen
  \bibfield  {author} {\bibinfo {author} {\bibfnamefont {E.}~\bibnamefont
  {Biham}}, \bibinfo {author} {\bibfnamefont {B.}~\bibnamefont {Huttner}}, \
  and\ \bibinfo {author} {\bibfnamefont {T.}~\bibnamefont {Mor}},\ }\href@noop
  {} {\bibfield  {journal} {\bibinfo  {journal} {Physical Review A}\ }\textbf
  {\bibinfo {volume} {54}},\ \bibinfo {pages} {2651} (\bibinfo {year}
  {1996})}\BibitemShut {NoStop}%
\bibitem [{\citenamefont {Inamori}(2002)}]{inamori2002}%
  \BibitemOpen
  \bibfield  {author} {\bibinfo {author} {\bibfnamefont {H.}~\bibnamefont
  {Inamori}},\ }\href@noop {} {\bibfield  {journal} {\bibinfo  {journal}
  {Algorithmica}\ }\textbf {\bibinfo {volume} {34}},\ \bibinfo {pages} {340}
  (\bibinfo {year} {2002})}\BibitemShut {NoStop}%
\bibitem [{\citenamefont {Ma}\ \emph {et~al.}(2012)\citenamefont {Ma},
  \citenamefont {Fung},\ and\ \citenamefont {Razavi}}]{Ma2012}%
  \BibitemOpen
  \bibfield  {author} {\bibinfo {author} {\bibfnamefont {X.}~\bibnamefont
  {Ma}}, \bibinfo {author} {\bibfnamefont {C.-H.~F.}\ \bibnamefont {Fung}}, \
  and\ \bibinfo {author} {\bibfnamefont {M.}~\bibnamefont {Razavi}},\ }\href
  {\doibase 10.1103/PhysRevA.86.052305} {\bibfield  {journal} {\bibinfo
  {journal} {Physical Review A}\ }\textbf {\bibinfo {volume} {86}} (\bibinfo
  {year} {2012}),\ 10.1103/PhysRevA.86.052305},\ \Eprint
  {http://arxiv.org/abs/1210.3929} {arXiv:1210.3929 [quant-ph]} \BibitemShut
  {NoStop}%
\bibitem [{\citenamefont {Kaneda}\ \emph {et~al.}(2017)\citenamefont {Kaneda},
  \citenamefont {Xu}, \citenamefont {Chapman},\ and\ \citenamefont
  {Kwiat}}]{kaneda2017}%
  \BibitemOpen
  \bibfield  {author} {\bibinfo {author} {\bibfnamefont {F.}~\bibnamefont
  {Kaneda}}, \bibinfo {author} {\bibfnamefont {F.}~\bibnamefont {Xu}}, \bibinfo
  {author} {\bibfnamefont {J.}~\bibnamefont {Chapman}}, \ and\ \bibinfo
  {author} {\bibfnamefont {P.~G.}\ \bibnamefont {Kwiat}},\ }\href@noop {}
  {\bibfield  {journal} {\bibinfo  {journal} {Optica}\ }\textbf {\bibinfo
  {volume} {4}},\ \bibinfo {pages} {1034} (\bibinfo {year} {2017})}\BibitemShut
  {NoStop}%
\bibitem [{\citenamefont {Liu}\ \emph {et~al.}(2013)\citenamefont {Liu},
  \citenamefont {Chen}, \citenamefont {Wang}, \citenamefont {Liang},
  \citenamefont {Shentu}, \citenamefont {Wang}, \citenamefont {Cui},
  \citenamefont {Yin}, \citenamefont {Liu}, \citenamefont {Li} \emph
  {et~al.}}]{liu2013}%
  \BibitemOpen
  \bibfield  {author} {\bibinfo {author} {\bibfnamefont {Y.}~\bibnamefont
  {Liu}}, \bibinfo {author} {\bibfnamefont {T.-Y.}\ \bibnamefont {Chen}},
  \bibinfo {author} {\bibfnamefont {L.-J.}\ \bibnamefont {Wang}}, \bibinfo
  {author} {\bibfnamefont {H.}~\bibnamefont {Liang}}, \bibinfo {author}
  {\bibfnamefont {G.-L.}\ \bibnamefont {Shentu}}, \bibinfo {author}
  {\bibfnamefont {J.}~\bibnamefont {Wang}}, \bibinfo {author} {\bibfnamefont
  {K.}~\bibnamefont {Cui}}, \bibinfo {author} {\bibfnamefont {H.-L.}\
  \bibnamefont {Yin}}, \bibinfo {author} {\bibfnamefont {N.-L.}\ \bibnamefont
  {Liu}}, \bibinfo {author} {\bibfnamefont {L.}~\bibnamefont {Li}},  \emph
  {et~al.},\ }\href@noop {} {\bibfield  {journal} {\bibinfo  {journal}
  {Physical Review Letters}\ }\textbf {\bibinfo {volume} {111}},\ \bibinfo
  {pages} {130502} (\bibinfo {year} {2013})}\BibitemShut {NoStop}%
\bibitem [{\citenamefont {Pirandola}\ \emph {et~al.}(2015)\citenamefont
  {Pirandola}, \citenamefont {Ottaviani}, \citenamefont {Spedalieri},
  \citenamefont {Weedbrook}, \citenamefont {Braunstein}, \citenamefont {Lloyd},
  \citenamefont {Gehring}, \citenamefont {Jacobsen},\ and\ \citenamefont
  {Andersen}}]{pirandola2015}%
  \BibitemOpen
  \bibfield  {author} {\bibinfo {author} {\bibfnamefont {S.}~\bibnamefont
  {Pirandola}}, \bibinfo {author} {\bibfnamefont {C.}~\bibnamefont
  {Ottaviani}}, \bibinfo {author} {\bibfnamefont {G.}~\bibnamefont
  {Spedalieri}}, \bibinfo {author} {\bibfnamefont {C.}~\bibnamefont
  {Weedbrook}}, \bibinfo {author} {\bibfnamefont {S.~L.}\ \bibnamefont
  {Braunstein}}, \bibinfo {author} {\bibfnamefont {S.}~\bibnamefont {Lloyd}},
  \bibinfo {author} {\bibfnamefont {T.}~\bibnamefont {Gehring}}, \bibinfo
  {author} {\bibfnamefont {C.~S.}\ \bibnamefont {Jacobsen}}, \ and\ \bibinfo
  {author} {\bibfnamefont {U.~L.}\ \bibnamefont {Andersen}},\ }\href@noop {}
  {\bibfield  {journal} {\bibinfo  {journal} {Nature Photonics}\ }\textbf
  {\bibinfo {volume} {9}},\ \bibinfo {pages} {397} (\bibinfo {year}
  {2015})}\BibitemShut {NoStop}%
\bibitem [{\citenamefont {Da~Silva}\ \emph {et~al.}(2013)\citenamefont
  {Da~Silva}, \citenamefont {Vitoreti}, \citenamefont {Xavier}, \citenamefont
  {Do~Amaral}, \citenamefont {Tempor{\~a}o},\ and\ \citenamefont {Von
  Der~Weid}}]{da2013}%
  \BibitemOpen
  \bibfield  {author} {\bibinfo {author} {\bibfnamefont {T.~F.}\ \bibnamefont
  {Da~Silva}}, \bibinfo {author} {\bibfnamefont {D.}~\bibnamefont {Vitoreti}},
  \bibinfo {author} {\bibfnamefont {G.}~\bibnamefont {Xavier}}, \bibinfo
  {author} {\bibfnamefont {G.}~\bibnamefont {Do~Amaral}}, \bibinfo {author}
  {\bibfnamefont {G.}~\bibnamefont {Tempor{\~a}o}}, \ and\ \bibinfo {author}
  {\bibfnamefont {J.}~\bibnamefont {Von Der~Weid}},\ }\href@noop {} {\bibfield
  {journal} {\bibinfo  {journal} {Physical Review A}\ }\textbf {\bibinfo
  {volume} {88}},\ \bibinfo {pages} {052303} (\bibinfo {year}
  {2013})}\BibitemShut {NoStop}%
\bibitem [{\citenamefont {Tang}\ \emph
  {et~al.}(2014{\natexlab{b}})\citenamefont {Tang}, \citenamefont {Liao},
  \citenamefont {Xu}, \citenamefont {Qi}, \citenamefont {Qian},\ and\
  \citenamefont {Lo}}]{Tang2014}%
  \BibitemOpen
  \bibfield  {author} {\bibinfo {author} {\bibfnamefont {Z.}~\bibnamefont
  {Tang}}, \bibinfo {author} {\bibfnamefont {Z.}~\bibnamefont {Liao}}, \bibinfo
  {author} {\bibfnamefont {F.}~\bibnamefont {Xu}}, \bibinfo {author}
  {\bibfnamefont {B.}~\bibnamefont {Qi}}, \bibinfo {author} {\bibfnamefont
  {L.}~\bibnamefont {Qian}}, \ and\ \bibinfo {author} {\bibfnamefont {H.-K.}\
  \bibnamefont {Lo}},\ }\href {\doibase 10.1103/PhysRevLett.112.190503}
  {\bibfield  {journal} {\bibinfo  {journal} {Phys. Rev. Lett.}\ }\textbf
  {\bibinfo {volume} {112}},\ \bibinfo {pages} {190503} (\bibinfo {year}
  {2014}{\natexlab{b}})}\BibitemShut {NoStop}%
\bibitem [{\citenamefont {Hwang}(2003)}]{Hwang2003}%
  \BibitemOpen
  \bibfield  {author} {\bibinfo {author} {\bibfnamefont {W.~Y.}\ \bibnamefont
  {Hwang}},\ }\href {\doibase 10.1103/PhysRevLett.91.057901} {\bibfield
  {journal} {\bibinfo  {journal} {Physical Review Letters 91, 057901}\ }
  (\bibinfo {year} {2003}),\ 10.1103/PhysRevLett.91.057901},\ \Eprint
  {http://arxiv.org/abs/quant-ph/0211153} {arXiv:quant-ph/0211153 [quant-ph]}
  \BibitemShut {NoStop}%
\bibitem [{\citenamefont {Lo}\ \emph {et~al.}(2005)\citenamefont {Lo},
  \citenamefont {Ma},\ and\ \citenamefont {Chen}}]{Lo2005}%
  \BibitemOpen
  \bibfield  {author} {\bibinfo {author} {\bibfnamefont {H.-K.}\ \bibnamefont
  {Lo}}, \bibinfo {author} {\bibfnamefont {X.}~\bibnamefont {Ma}}, \ and\
  \bibinfo {author} {\bibfnamefont {K.}~\bibnamefont {Chen}},\ }\href {\doibase
  10.1103/PhysRevLett.94.230504} {\bibfield  {journal} {\bibinfo  {journal}
  {Phys. Rev. Lett.}\ }\textbf {\bibinfo {volume} {94}},\ \bibinfo {pages}
  {230504} (\bibinfo {year} {2005})}\BibitemShut {NoStop}%
\bibitem [{\citenamefont {Wang}(2005)}]{wang2005beating}%
  \BibitemOpen
  \bibfield  {author} {\bibinfo {author} {\bibfnamefont {X.-B.}\ \bibnamefont
  {Wang}},\ }\href@noop {} {\bibfield  {journal} {\bibinfo  {journal} {Physical
  review letters}\ }\textbf {\bibinfo {volume} {94}},\ \bibinfo {pages}
  {230503} (\bibinfo {year} {2005})}\BibitemShut {NoStop}%
\bibitem [{\citenamefont {Yu}\ \emph {et~al.}(2015)\citenamefont {Yu},
  \citenamefont {Zhou},\ and\ \citenamefont {Wang}}]{yu2015}%
  \BibitemOpen
  \bibfield  {author} {\bibinfo {author} {\bibfnamefont {Z.-W.}\ \bibnamefont
  {Yu}}, \bibinfo {author} {\bibfnamefont {Y.-H.}\ \bibnamefont {Zhou}}, \ and\
  \bibinfo {author} {\bibfnamefont {X.-B.}\ \bibnamefont {Wang}},\ }\href@noop
  {} {\bibfield  {journal} {\bibinfo  {journal} {Physical Review A}\ }\textbf
  {\bibinfo {volume} {91}},\ \bibinfo {pages} {032318} (\bibinfo {year}
  {2015})}\BibitemShut {NoStop}%
\bibitem [{\citenamefont {Zhou}\ \emph {et~al.}(2016)\citenamefont {Zhou},
  \citenamefont {Yu},\ and\ \citenamefont {Wang}}]{Zhou2016}%
  \BibitemOpen
  \bibfield  {author} {\bibinfo {author} {\bibfnamefont {Y.-H.}\ \bibnamefont
  {Zhou}}, \bibinfo {author} {\bibfnamefont {Z.-W.}\ \bibnamefont {Yu}}, \ and\
  \bibinfo {author} {\bibfnamefont {X.-B.}\ \bibnamefont {Wang}},\ }\href@noop
  {} {\bibfield  {journal} {\bibinfo  {journal} {Physical Review A}\ }\textbf
  {\bibinfo {volume} {93}},\ \bibinfo {pages} {42324} (\bibinfo {year}
  {2016})}\BibitemShut {NoStop}%
\bibitem [{\citenamefont {Ghassemlooy}\ \emph {et~al.}(2012)\citenamefont
  {Ghassemlooy}, \citenamefont {Popoola},\ and\ \citenamefont
  {Rajbhandari}}]{Ghassemlooy2012}%
  \BibitemOpen
  \bibfield  {author} {\bibinfo {author} {\bibfnamefont {Z.}~\bibnamefont
  {Ghassemlooy}}, \bibinfo {author} {\bibfnamefont {W.}~\bibnamefont
  {Popoola}}, \ and\ \bibinfo {author} {\bibfnamefont {S.}~\bibnamefont
  {Rajbhandari}},\ }\href@noop {} {\emph {\bibinfo {title} {Optical wireless
  communications: system and channel modelling with Matlab{\textregistered}}}}\
  (\bibinfo  {publisher} {CRC press},\ \bibinfo {year} {2012})\BibitemShut
  {NoStop}%
\bibitem [{\citenamefont {Goodman}(2015)}]{Goodman2015}%
  \BibitemOpen
  \bibfield  {author} {\bibinfo {author} {\bibfnamefont {J.~W.}\ \bibnamefont
  {Goodman}},\ }\href@noop {} {\emph {\bibinfo {title} {Statistical optics}}}\
  (\bibinfo  {publisher} {John Wiley \& Sons},\ \bibinfo {year}
  {2015})\BibitemShut {NoStop}%
\bibitem [{\citenamefont {Karp}\ \emph {et~al.}(2013)\citenamefont {Karp},
  \citenamefont {Gagliardi}, \citenamefont {Moran},\ and\ \citenamefont
  {Stotts}}]{Karp2013}%
  \BibitemOpen
  \bibfield  {author} {\bibinfo {author} {\bibfnamefont {S.}~\bibnamefont
  {Karp}}, \bibinfo {author} {\bibfnamefont {R.~M.}\ \bibnamefont {Gagliardi}},
  \bibinfo {author} {\bibfnamefont {S.~E.}\ \bibnamefont {Moran}}, \ and\
  \bibinfo {author} {\bibfnamefont {L.~B.}\ \bibnamefont {Stotts}},\
  }\href@noop {} {\emph {\bibinfo {title} {Optical channels: fibers, clouds,
  water, and the atmosphere}}}\ (\bibinfo  {publisher} {Springer Science \&
  Business Media},\ \bibinfo {year} {2013})\BibitemShut {NoStop}%
\bibitem [{\citenamefont {Berk}\ \emph {et~al.}(1987)\citenamefont {Berk},
  \citenamefont {Bernstein},\ and\ \citenamefont
  {Robertson}}]{berk1987modtran}%
  \BibitemOpen
  \bibfield  {author} {\bibinfo {author} {\bibfnamefont {A.}~\bibnamefont
  {Berk}}, \bibinfo {author} {\bibfnamefont {L.~S.}\ \bibnamefont {Bernstein}},
  \ and\ \bibinfo {author} {\bibfnamefont {D.~C.}\ \bibnamefont {Robertson}},\
  }\href@noop {} {\emph {\bibinfo {title} {MODTRAN: A moderate resolution model
  for LOWTRAN}}},\ \bibinfo {type} {Tech. Rep.}\ \bibinfo {number}
  {SSI-TR-124}\ (\bibinfo  {institution} {Spectral Sciences Inc., Burlington
  MA},\ \bibinfo {address} {Burlington, MA},\ \bibinfo {year}
  {1987})\BibitemShut {NoStop}%
\bibitem [{\citenamefont {Berk}\ \emph {et~al.}(2016)\citenamefont {Berk},
  \citenamefont {van~den Bosch}, \citenamefont {Hawes}, \citenamefont
  {Perkins}, \citenamefont {Conforti}, \citenamefont {Anderson}, \citenamefont
  {Kennett},\ and\ \citenamefont {Acharya}}]{Berk2016}%
  \BibitemOpen
  \bibfield  {author} {\bibinfo {author} {\bibfnamefont {A.}~\bibnamefont
  {Berk}}, \bibinfo {author} {\bibfnamefont {J.}~\bibnamefont {van~den Bosch}},
  \bibinfo {author} {\bibfnamefont {F.}~\bibnamefont {Hawes}}, \bibinfo
  {author} {\bibfnamefont {T.}~\bibnamefont {Perkins}}, \bibinfo {author}
  {\bibfnamefont {P.~F.}\ \bibnamefont {Conforti}}, \bibinfo {author}
  {\bibfnamefont {G.~P.}\ \bibnamefont {Anderson}}, \bibinfo {author}
  {\bibfnamefont {R.~G.}\ \bibnamefont {Kennett}}, \ and\ \bibinfo {author}
  {\bibfnamefont {P.~K.}\ \bibnamefont {Acharya}},\ }\href@noop {} {\emph
  {\bibinfo {title} {MODTRAN\textregistered 6.0.0 (Revision 5) User’s
  Manual}}} (\bibinfo {year} {2016})\BibitemShut {NoStop}%
\bibitem [{\citenamefont {Smith}\ \emph {et~al.}(1978)\citenamefont {Smith},
  \citenamefont {Dube}, \citenamefont {Gardner}, \citenamefont {Clough},\ and\
  \citenamefont {Kneizys}}]{smith1978fascode}%
  \BibitemOpen
  \bibfield  {author} {\bibinfo {author} {\bibfnamefont {H.}~\bibnamefont
  {Smith}}, \bibinfo {author} {\bibfnamefont {D.}~\bibnamefont {Dube}},
  \bibinfo {author} {\bibfnamefont {M.}~\bibnamefont {Gardner}}, \bibinfo
  {author} {\bibfnamefont {S.}~\bibnamefont {Clough}}, \ and\ \bibinfo {author}
  {\bibfnamefont {F.}~\bibnamefont {Kneizys}},\ }\href@noop {} {\emph {\bibinfo
  {title} {FASCODE-fast atmospheric signature code (spectral transmittance and
  radiance)}}},\ \bibinfo {type} {Tech. Rep.}\ (\bibinfo  {institution}
  {VISIDYNE INC BURLINGTON MA},\ \bibinfo {year} {1978})\BibitemShut {NoStop}%
\bibitem [{\citenamefont {Comandar}\ \emph {et~al.}(2016)\citenamefont
  {Comandar}, \citenamefont {Lucamarini}, \citenamefont {Fröhlich},
  \citenamefont {Dynes}, \citenamefont {Sharpe}, \citenamefont {Tam},
  \citenamefont {Yuan}, \citenamefont {Penty},\ and\ \citenamefont
  {Shields}}]{Comandar2016}%
  \BibitemOpen
  \bibfield  {author} {\bibinfo {author} {\bibfnamefont {L.~C.}\ \bibnamefont
  {Comandar}}, \bibinfo {author} {\bibfnamefont {M.}~\bibnamefont
  {Lucamarini}}, \bibinfo {author} {\bibfnamefont {B.}~\bibnamefont
  {Fröhlich}}, \bibinfo {author} {\bibfnamefont {J.~F.}\ \bibnamefont
  {Dynes}}, \bibinfo {author} {\bibfnamefont {A.~W.}\ \bibnamefont {Sharpe}},
  \bibinfo {author} {\bibfnamefont {S.~W.-B.}\ \bibnamefont {Tam}}, \bibinfo
  {author} {\bibfnamefont {Z.~L.}\ \bibnamefont {Yuan}}, \bibinfo {author}
  {\bibfnamefont {R.~V.}\ \bibnamefont {Penty}}, \ and\ \bibinfo {author}
  {\bibfnamefont {A.~J.}\ \bibnamefont {Shields}},\ }\href {\doibase
  10.1038/nphoton.2016.50} {\bibfield  {journal} {\bibinfo  {journal} {Nature
  Photonics}\ }\textbf {\bibinfo {volume} {10}},\ \bibinfo {pages} {312}
  (\bibinfo {year} {2016})}\BibitemShut {NoStop}%
\bibitem [{\citenamefont {Wang}\ \emph {et~al.}(2015)\citenamefont {Wang},
  \citenamefont {Li}, \citenamefont {Wu}, \citenamefont {Liu}, \citenamefont
  {Hu},\ and\ \citenamefont {Xu}}]{Wang2015}%
  \BibitemOpen
  \bibfield  {author} {\bibinfo {author} {\bibfnamefont {H.}~\bibnamefont
  {Wang}}, \bibinfo {author} {\bibfnamefont {B.}~\bibnamefont {Li}}, \bibinfo
  {author} {\bibfnamefont {X.}~\bibnamefont {Wu}}, \bibinfo {author}
  {\bibfnamefont {C.}~\bibnamefont {Liu}}, \bibinfo {author} {\bibfnamefont
  {Z.}~\bibnamefont {Hu}}, \ and\ \bibinfo {author} {\bibfnamefont
  {P.}~\bibnamefont {Xu}},\ }\href {\doibase 10.1080/09500340.2015.1037801}
  {\bibfield  {journal} {\bibinfo  {journal} {Optik - International Journal for
  Light and Electron Optics}\ }\textbf {\bibinfo {volume} {126}},\ \bibinfo
  {pages} {2726} (\bibinfo {year} {2015})}\BibitemShut {NoStop}%
\bibitem [{\citenamefont {Sprung}\ \emph {et~al.}(2014)\citenamefont {Sprung},
  \citenamefont {Sucher}, \citenamefont {Ramkilowan},\ and\ \citenamefont
  {Griffith}}]{Sprung2014}%
  \BibitemOpen
  \bibfield  {author} {\bibinfo {author} {\bibfnamefont {D.}~\bibnamefont
  {Sprung}}, \bibinfo {author} {\bibfnamefont {E.}~\bibnamefont {Sucher}},
  \bibinfo {author} {\bibfnamefont {A.}~\bibnamefont {Ramkilowan}}, \ and\
  \bibinfo {author} {\bibfnamefont {D.~J.}\ \bibnamefont {Griffith}},\ }in\
  \href {\doibase 10.1117/12.2070257} {\emph {\bibinfo {booktitle} {Remote
  Sensing of Clouds and the Atmosphere XIX; and Optics in Atmospheric
  Propagation and Adaptive Systems XVII}}},\ \bibinfo {series} {Society of
  Photo-Optical Instrumentation Engineers (SPIE) Conference Series}, Vol.\
  \bibinfo {volume} {9242},\ \bibinfo {editor} {edited by\ \bibinfo {editor}
  {\bibfnamefont {A.}~\bibnamefont {{Comern}}}, \bibinfo {editor}
  {\bibfnamefont {K.}~\bibnamefont {{Stein}}}, \bibinfo {editor} {\bibfnamefont
  {E.~I.}\ \bibnamefont {{Kassianov}}}, \bibinfo {editor} {\bibfnamefont
  {J.~D.}\ \bibnamefont {{Gonglewski}}}, \ and\ \bibinfo {editor}
  {\bibfnamefont {K.}~\bibnamefont {{Schfer}}}}\ (\bibinfo {year} {2014})\ p.\
  \bibinfo {pages} {92421I}\BibitemShut {NoStop}%
\bibitem [{\citenamefont {Qing}\ \emph {et~al.}(2018)\citenamefont {Qing},
  \citenamefont {Wu}, \citenamefont {Li}, \citenamefont {Tian}, \citenamefont
  {Liu}, \citenamefont {Rao},\ and\ \citenamefont {Zhu}}]{Qing2018}%
  \BibitemOpen
  \bibfield  {author} {\bibinfo {author} {\bibfnamefont {C.}~\bibnamefont
  {Qing}}, \bibinfo {author} {\bibfnamefont {X.}~\bibnamefont {Wu}}, \bibinfo
  {author} {\bibfnamefont {X.}~\bibnamefont {Li}}, \bibinfo {author}
  {\bibfnamefont {Q.}~\bibnamefont {Tian}}, \bibinfo {author} {\bibfnamefont
  {D.}~\bibnamefont {Liu}}, \bibinfo {author} {\bibfnamefont {R.}~\bibnamefont
  {Rao}}, \ and\ \bibinfo {author} {\bibfnamefont {W.}~\bibnamefont {Zhu}},\
  }\href {\doibase 10.3847/1538-3881/aa9e8f} {\bibfield  {journal} {\bibinfo
  {journal} {The Astronomical Journal}\ }\textbf {\bibinfo {volume} {155}},\
  \bibinfo {pages} {37} (\bibinfo {year} {2018})}\BibitemShut {NoStop}%
\end{thebibliography}%

\begin{appendices}
\section{Noise Model and Finite Key Calculation}\label{keyCalculation}

Here we describe the model needed to predict the sifted key and error quantities, then proceed to compute the finite secure key length using a method described in an appendix of Ref.\ \cite{Wang2019a}. We consider intensities of Alice's and Bob's beams in the set $\{ s_{A,B}, \mu_{A,B}, \nu_{A,B}, \omega \}$, where $s$ is the signal, $\mu$ and $\nu$ are decoy states, and $\omega =0$ is the vacuum state, assuming perfect intensity modulators. $\eta_D$ is the detector efficiency for all detectors and $\eta_{A,B}$ are the channel transmittances excluding the detector. 

\subsection{Noise Model}

\subsubsection{Z Basis}

The probability of coincidence when Alice and Bob send opposite polarizations is
\begin{align}
	n_{z1} = \frac{1}{2}(1 - 2e_{d,Z})(1 - \textrm{e}^{-\eta_A\eta_Ds_A})(1 - \textrm{e}^{-\eta_B\eta_Ds_B})
\end{align}
where dark counts are neglected, as well as cases of Alice and Bob both being misaligned. $P_{s_{A,B}}$ and $N$ are suppressed in this step, because they are canceled in the secure key calculation.

Whenever Alice and Bob send the same polarization, the coincident probability is
\begin{multline}
	n_{z2} = \frac{1}{2}(1 - \textrm{e}^{-(1-e_{d,Z})\eta_A\eta_Ds_A}\textrm{e}^{-(1-e_{d,Z})\eta_B\eta_Ds_B})*\\*(e_{d,Z}\eta_A\eta_Ds_A + e_{d,Z}\eta_B\eta_Ds_B + 2Y_0)
\end{multline}
We have $n_Z = n_{z1} + n_{z2}$ and $m_X = n_{z2}$.

\subsubsection{X Basis}

We will refer to each intensity as $k_i$, where $i=2, 3$ are the decoy states, and $i=4$ is the vacuum. The only coincidences that survive sifting have the same intensity state. 

First, consider when Alice and Bob send opposite polarizations. Whenever a single photon is incident at the beam splitter, the only coincidences that are possible are due to dark counts. We have
\begin{multline}
	P_{coin} = \eta_A\eta_Dk_{i,A}\textrm{e}^{-\eta_A\eta_D{k_{i,A}}}Y_0 +\\+ \eta_B\eta_Dk_{j,B}\textrm{e}^{-\eta_B\eta_Dk_{j,B}}Y_0
\end{multline}
When Alice and Bob each send one photon, we have
\begin{align}
	P_{coin} = \eta_A\eta_B\eta_D^2k_{i,A}k_{j,B}\textrm{e}^{-\eta_A\eta_D{k_{i,A}} - \eta_B\eta_D{k_{j,B}}}
\end{align} 
In the case where Alice or Bob sends 2 photons and the other sends no photons, we have
\begin{align}
	\eta_A\eta_B\eta_D^2\textrm{e}^{-\eta_A\eta_D{k_{i,A}} - \eta_B\eta_D{k_{j,B}}}\frac{k_{i,A}^2 + k_{j,B}^2}{2}
\end{align}
Three-photon events are not considered, so the model loses accuracy at lower losses and high photon numbers. 

The number of $\ket{\psi^-}$ events is
\begin{multline}
	n_{c1} = \frac{1}{2}((\eta_A\eta_Dk_{i,A}\textrm{e}^{-\eta_A\eta_D{k_{i,A}}}Y_0 +\\ \eta_B\eta_Dk_{j,B}\textrm{e}^{-\eta_B\eta_Dk_{j,B}}Y_0) +\\ .5*(\eta_A\eta_B\eta_D^2k_{i,A}k_{j,B}\textrm{e}^{-\eta_A\eta_D{k_{i,A}} - \eta_B\eta_D{k_{j,B}}})(1-2e_{d,X}) +\\ .25*(\eta_A\eta_B\eta_D^2\textrm{e}^{-\eta_A\eta_D{k_{i,A}} - \eta_B\eta_D{k_{j,B}}}*\\\frac{k_{i,A}^2 + k_{j,B}^2}{2}))
\end{multline} 
and the number of $\ket{\psi^+}$ events is:
\begin{multline}
	n_{w1} = \frac{1}{2}((\eta_A\eta_D k_{i,A}\textrm{e}^{-\eta_A\eta_D{k_{i,A}}}Y_0 +\\ \eta_B\eta_D k_{j,B}\textrm{e}^{-\eta_B\eta_D k_{j,B}}Y_0) +\\ (\eta_A\eta_B\eta_D^2k_{i,A}k_{j,B}\textrm{e}^{-\eta_A\eta_D{k_{i,A}} - \eta_B\eta_D{k_{j,B}}})e_{d,X} +\\ .25\eta_A\eta_B\eta_D^2\textrm{e}^{-\eta_A\eta_D{k_{i,A}} - \eta_B\eta_D{k_{j,B}}}\frac{k_{i,A}^2 + k_{j,B}^2}{2})
\end{multline} 
When Alice and Bob send the same polarization state, the analysis is similar, except with the roles of $\ket{\psi^+}$ and $\ket{\psi^-}$ exchanged. Therefore, $n_{c2} = n_{c1}$ and $n_{w2} = n_{w1}$. Finally, we have
\begin{align}
	n_X^{i,j} &= 2(n_{c1} + n_{w1}) \nonumber\\
	m_X^{i,j} &= 2n_{w1}
\end{align}

\subsection{Secure Key Calculation}

The set of probabilities $\{ n_Z, n_X, m_Z, m_X \}$ derived above can be used to calculate the secure key rate, by mostly following the steps in the appendix of Ref.\ \cite{Wang2019a}. Having suppressed $N, P_{k_{i,j}}$, the gains are given by
\begin{align}
	Q_X^{i,j} &= n_X^{i,j} \nonumber\\
	T_X^{i,j} &= m_X^{i,j}
\end{align}
where $n_X^{i,j}$ applies to all decoy intensities $i\in \{ 2,3,4\}$. Next, we apply bounds according to $\gamma = 5.3$, the number of standard deviations of an observed value from the expected. This value of $\gamma$ corresponds to a failure probability of less than $10^{-7}$.  We have
\begin{align*}
	\overline{Q_X^{i,j}} = Q_X^{i,j} + \gamma\sqrt{\frac{Q_X^{i,j} }{NP_{k_i}P_{k_j}}}\\
	\underline{Q_X^{i,j}} = Q_X^{i,j} - \gamma\sqrt{\frac{Q_X^{i,j} }{NP_{k_i}P_{k_j}}}\\
	\overline{T_X^{i,j}} = T_X^{i,j} + \gamma\sqrt{\frac{T_X^{i,j} }{NP_{k_i}P_{k_j}}}\\
	\underline{T_X^{i,j}} = T_X^{i,j} - \gamma\sqrt{\frac{T_X^{i,j} }{NP_{k_i}P_{k_j}}}
\end{align*}
Then, we define
\begin{align*}
	\underline{Q_{M1}^{\nu\nu}} = \textrm{e}^{\nu_A + \nu_B}\underline{Q_X^{\nu\nu}} - \textrm{e}^{\nu_A}\overline{Q_X^{\nu\omega}} - \textrm{e}^{\nu_B}\overline{Q_X^{\omega\nu}} + \underline{Q_X^{\omega\omega}}\\
	\overline{Q_{M2}^{\mu\mu}} = \textrm{e}^{\mu_A + \mu_B}\overline{Q_X^{\mu\mu}} - \textrm{e}^{\mu_A}\underline{Q_X^{\mu\omega}} - \textrm{e}^{\mu_B}\underline{Q_X^{\omega\mu}} + \underline{Q_X^{\omega\omega}}
\end{align*}
We place lower and upper bounds on yield,
\begin{align}
	Y_{X,min}^{1,1} = \frac{1}{\mu_A-\nu_A}(\frac{\mu_A}{\nu_A\nu_B}\underline{Q_{M1}^{\nu\nu}} - \frac{\nu_A}{\mu_A\mu_B}\overline{Q_{M2}^{\mu\mu}})
\end{align}
and error,
\begin{multline}
	e_{X,max}^{1,1} = \frac{1}{\nu_A\nu_BY_{X,min}^{1,1}}(\textrm{e}^{\nu_A + \nu_B}\overline{T_{\nu\nu}} - \textrm{e}^{\nu_A}\underline{T_{\nu\omega}} -\\ \textrm{e}^{\nu_B}\underline{T_{\omega\nu}} + \overline{T_{\omega\omega}})
\end{multline}
Finally, we define $E_z$ as the error rate in the $Z$ basis. The secure key rate is
\begin{multline}
	R = P_{s_A}P_{s_B}(s_As_B\textrm{e}^{-(s_A+s_B)}Y_{X,min}^{1,1}[1 - h_2(e_{X,max}^{1,1})] -\\ f_{EC}Q_{Z}^{1,1}H_2(E_z))
\end{multline}
where $f_{EC}$ is the error correcting efficiency which we set to 1.16, and $H_2$ is the binary Shannon entropy. 

\end{appendices}

\end{document}